\newif\ifMNRAS
\MNRASfalse 

\PassOptionsToPackage{pdfpagelabels=false}{hyperref} 

\ifMNRAS
    \documentclass[fleqn,usenatbib,useAMS]{mnras}
    \usepackage{graphicx}
\else

\documentclass[nolinenumber]{aastex63}
\fi

\usepackage{acronym}
\usepackage{hyperref}
\usepackage{amsmath,amsfonts,amssymb}
\usepackage{mathrsfs}
\usepackage{mathtools}
\usepackage[utf8]{inputenc}

\usepackage[normalem]{ulem}

\newcommand{\tianshu}[1]{{#1}}
\newcommand{\adam}[1]{{#1}}

\newcommand{\equ}[1]{
\begin{equation}
\begin{split}
#1
\end{split}
\end{equation}
}

\date{Received February 28, 2025}

\shorttitle{Fast-Flavor Conversion in CCSNe}

\begin{document}
\title{The Effect of the Fast-Flavor Instability on Core-Collapse Supernova Models} 
\correspondingauthor{Tianshu Wang}
\email{tianshuw@berkeley.edu}
\author[0000-0002-0042-9873]{Tianshu Wang}
\affiliation{Department of Physics, University of California, Berkeley, CA, 94720-7300 USA}
\author[0000-0002-3099-5024]{Adam Burrows}
\affiliation{Department of Astrophysical Sciences, Princeton University, NJ 08544, USA; School of Natural Sciences, Institute for Advanced Study, Princeton, NJ 08540}
   
\begin{abstract}
Merging our supernova code F{\sc{ornax}} with the Box3D fast-flavor neutrino oscillation formalism, we explore the effects of fast-flavor conversion (FFC) in state-of-the-art 1D and 2D core-collapse supernova simulations. We find that after a few tens of milliseconds after bounce the FFC emerges just interior to and exterior to the stalled shock wave. It does not obtain in the PNS core nor near the average neutrinosphere radii. Interior to the shock, this results in a temporary change in the net neutrino heating rate of $\sim$10\%, due mostly to a hardening of the $\nu_e$ and $\bar{\nu}_e$ neutrino spectra, despite the decrease in their corresponding neutrino number fluxes. In 1D, the hydrodynamic effects are not large, with increases in the stalled shock radius by of order ten to twenty kilometers that abate within a few hundred milliseconds. In 2D, the hydrodynamic effect of the FFC is a bit more noticeable, resulting in slightly earlier explosions for models for lower-mass progenitors, but also potentially inhibiting explosions for some higher-mass progenitors. Fast-flavor conversion continues to operate at larger radii at later times. The net result is a shift upward in the $\nu_{\mu}$ energy and number luminosities and a shift downward in the same quantities for both the $\nu_e$ and $\bar{\nu}_e$ neutrinos. There seems to be a trend at very large radii and later times towards partial species and spectral equipartition. If this is true, it could be an interesting feature of supernova neutrino detection at later times in underground and under-ice facilities.  
\end{abstract} 

\ifMNRAS
\else 
\fi

\section{Introduction}
\label{sec:int}  

Neutrinos have been considered the central driving agents of the mechanism of the supernova explosions of massive stars since the pioneering work in of \citet{CoWh66} and \citet{arnett1967} in the 1960's. The concept of the delay in the ``ignition" of the explosion after the bounce at nuclear densities of the Chandrasekhar core birthed in a massive star at its terminal phase was advanced by \citet{wilson1985} and \citet{bethe_wilson}. Though the posited necessary boost in the driving neutrino luminosity due to doubly-diffusive ``neutron fingers" was shown not to occur \citep{bruenn_dineva}, it was later demonstrated that neutrino-driven convection could facilitate explosion \citep{herant1994,burrows:95,janka1996}, in particular aided by the turbulent stress of neutrino-driven convection \citep{burrows:95}.  Since this foundation was laid, there has been an avalanche of investigations over the ensuing decades on all aspects of core collapse and explosion, with the result that the field has matured into a vigorous, if complicated, branch of astrophysics. Recent detailed three-dimensional radiation/hydrodynamic simulations \citep{lentz:15,burrows_2019,vartanyan2019,muller_lowmass,stockinger2020,burrows_2020,bollig2021,sandoval2021,nakamura2022,vartanyan2023,burrows_correlations_2024} have demonstrated the viability of the neutrino mechanism and the complexity of the phenomenon \citep{janka2012,burrows2013,burrows_vartanyan_nature,janka2025}.

However, in order to make progress, this very complexity has necessitated approximations, however modest, to the physical inputs and numerical algorithms.  One aspect of neutrino physics whose consequences for supernova explosions and their products has been hovering unresolved is neutrino flavor oscillation, the possibility that neutrinos of different flavors could mix or swap. Since neutrino-matter interactions and production are flavor- and energy-dependent, flavor mixing could change explosion hydrodynamics. Given that in every decade our understanding of supernova physics has co-evolved with our understanding of neutrino physics \citep{Burrows:2018nu}, it is natural that the potential effects of neutrino oscillations need to be understood.

One modality of neutrino oscillation that has recently come into sharp focus is {``fast-flavor conversion" (FFC)}, whereby fast neutrino flavor transformations occur when and where the angular distributions of $\nu_e$ and $\bar{\nu}_e$ cross {and cause the fast-flavor instability (FFI)}.  Such electron-lepton number (ELN) crossing was suggested in the seminal works of \citet{samuelsPhysRevD.48.1462} and \citet{Sawyer_2005,Sawyer_2016}, but it took several years for the community to realize its potential importance. Now, there is a rich literature on its physics \citep{Chakraborty_2016,johns2020,tamborra2021,padilla-gay_raffelt2022,richers_sen2022,morinaga2022,johns2025} and quantum kinetics (QKE) \citep{volpe2015,richers2019,nagakura2023,nagakura2023b,volpe2024,xiong2024}, as well as on an interesting collisional instability \citep{johns_collisional2023,johns2025}. 

{However, the inclusion of FFC in time-dependent self-consistent CCSN simulations has to date been far from satisfactory, with the result that the impact on CCSN explosion properties remains very uncertain.} 
The major reason is that multi-group, multi-species and {\it multi-angle} radiation/hydrodynamic simulations in three spatial dimensions and for all the density matrix elements, including the off-diagonal terms that capture the quantum coherence and mixing \citep{Strack_2005,volpe2015,richers2019,volpe2024}, have not been possible.  Angular-moment methods that require higher-order closures have been developed  \citep{Zhang_2013,johns2020,nagakura_johns2021,myers_richers_moments2022,froustey2024,kneller_QKE2024}, but even they have proven too resource-intensive to employ. Importantly, the inverse timescale for FFI is $\sim$$\sqrt{2} G_F n_L$, where $n_L$ is the background {neutrino and anti-neutrino} density and $G_F$ is the Fermi coupling constant, and this in the supernova core can be nanoseconds. Such a timescale is much smaller than the computational timesteps of modern supernova codes (typically microseconds) , making the inclusion of FFC and its effects quite difficult. There have been attempts to identify where in the supernova core angular crossings could occur
\citep{abbar2019,glas_osc_2020,morinaga2020,abbar2021,johns_nagakura2021,nagakura_johns2021}, some quite sophisticated \citep{richers2021,richers2021b},
but these have not proven definitive. {Moreover, the feedback on supernova hydrodynamics has generally not been included}. There have also been attempts to perform supernova simulations with approximate models of the effects of FFC \citep{ehring2023a,ehring_help_or_hinder2023,mori2025_ffc}. These generally make ad hoc assumptions about flavor mixing and where it occurs and too have fallen short. {These approaches also sensitively depend on a free parameter (the density threshold below which flavor equipartition is instantaneously achieved) whose value and appropriateness is quite uncertain.}

Where there has been progress is in developing \tianshu{approximate} schemes that capture the basic physics of angular crossing and FFC, while preserving the basic \tianshu{conservation laws}. These methods \citep{richers2022,xiong_mengru2023,george2024,richers_box3D.2024} predict the asymptotic states of fast neutrino-flavor conversion and provide survival probabilities that can be incorporated into modern supernova codes. \tianshu{They rely on assumptions such as the analytic functional form of the survival probabilities, and tests using global transport simulations \citep{xiong2025} other than local box simulations with periodic boundary conditions are still called for.} Also developed are potentially related methods that take a ``thermodynamic" or a ``maximum entropy" approach \citep{johns2023,froustey_maxent2024}\tianshu{, which also require further tests}. \tianshu{With these caveats in mind, these approximate methods greatly accelerate the calculation of the FFC, and provide a potential way to incorporate FFC into classic CCSN simulations in a self-consistent time-dependent way.} 

Nevertheless, none of the \tianshu{above} prescriptions has yet been incorporated into a sophisticated time-dependent supernova code. With this paper, we make the first attempt to do so and to derive the potential consequences of the {FFC} in supernova hydrodynamics and supernova theory. We employ the Box3D scheme of \citet{richers_box3D.2024} that seems adequately to preserve the \tianshu{conservation laws in} fast-flavor conversion in the context of QKE and to predict survival probabilities that can rather easily be incorporated into modern supernova simulations.    

First, in \S\ref{method} we discuss our computational approach, and how we have generalized our core-collapse supernova (CCSN) code F{\sc{ornax}} to include the Box3D fast-flavor conversion methodology. Then, in \S\ref{results}
we describe our results and explore the differences in the hydrodynamic and neutrino sectors due to the inclusion of fast-flavor oscillations. Finally, we summarize our findings in \S\ref{conclusions}.

\section{Method}
\label{method}

The sophisticated code (F{\sc{ornax}}) we use for this study has been described in detail in \citet{skinner2019}, \citet{burrows_40}, and in the appendix to \citet{vartanyan2019}. The neutrino-matter microphysics used in its classical physics sector can be found in \citet{2006NuPhA.777..356B} and \citet{2020PhRvD.102b3017W}. In the past, F{\sc{ornax}} has been used in a variety of papers on a broad spectrum of topics in core-collapse supernova theory \citep[e.g.,][]{radice2017b,burrows2018,vartanyan2018a,vartanyan2018b,vsg2018,burrows_2019,radice2019,hiroki_2019,vartanyan2019,Nagakura2020,vartanyan2020,burrows_2020,nagakura2021,burrows_vartanyan_nature,2022MNRAS.510.4689V,coleman,burrows_40,burrows_correlations_2024}. Here, we employ the SFHo nuclear equation of state \citep{2013ApJ...774...17S}, 1024 radial zones for the 1D simulations, and for the 2D simulations a grid of $1024\times128$ ($r\times\theta$).
We use for both sets of simulations twelve neutrino energy groups from 1 to 300 MeV for each of three species ($\nu_e$, $\bar{\nu}_e$, and ``$\nu_{\mu}$" [$\equiv$ $\nu_{\mu}+\bar{\nu}_{\mu}+\nu_{\tau}+\bar{\nu}_{\tau}$]).
The outer boundary is set at 30,000 kilometers (km) and the inner radial zone is 0.5 km wide. The progenitor models are taken from \citet{swbj16} and \citet{sukhbold2018}.

For this study, and to determine the effects of fast-flavor instability and conversion, we have added to our classical F{\sc{ornax}} code the Box3D oscillation formalism \citep{richers_box3D.2024}. {Based on the Box-like scheme proposed in \citet{zaizen2023}, the Box3D scheme is an approximate method for predicting asymptotic states of fast neutrino-flavor conversion, generalized to handle non-axisymmetric angular distributions.} Note that for most of the inner region the timescale for fast-flavor conversion is much shorter than the $\sim$microsecond timescale of the hydrodynamic timestep. With the survival probabilities provided we oscillate one neutrino species into another, while preserving the \tianshu{conservation laws in} fast-flavor conversions in the context of the quantum kinetic equations \citep{volpe2015,richers2019,volpe2024,kneller_QKE2024}. In this method, heavy-lepton flavor neutrinos are assumed to have equal distributions ($f_{{\nu}_\mu}=f_{{\nu}_\tau}=f_{{\nu}_x}$). The same assumption is been made for heavy-lepton flavor antineutrinos ($f_{\bar{\nu}_\mu}=f_{\bar{\nu}_\tau}=f_{\bar{\nu}_x}$). Given the neutrino distribution $f_{\nu_\alpha}(E,\hat{n})$ (where $E$ is the neutrino energy and $\hat{n}$ is the momentum direction vector), the $\alpha-$flavor lepton number angular distribution is (for simplicity we assume here $c=\hbar=1$):
\equ{
G_{\alpha}(\hat{n}) = \sqrt{2}G_F\int_{0}^{\infty}\frac{ {\rm d}E\,E^2}{(2\pi)^3}(f_{\nu_\alpha}(E,\hat{n})-f_{\bar{\nu}_\alpha}(E,\hat{n}))\, ,
}
and the ELN-XLN distribution in a given direction is
\equ{
G_{\hat{n}} = G_{e}(\hat{n}) - G_{x}(\hat{n})\, .
}

The sign of $G$ divides the angular space into two regions: $\Gamma_+=\{\hat{n}\vert G(\hat{n})>0\}$ and $\Gamma_-=\{\hat{n}\vert G(\hat{n})<0\}$. We define integrals on these two regions:
\equ{
I_+=&\int_{\Gamma_+} {\rm d}\hat{n}G_{\hat{n}},\,\,\,\,I_-=-\int_{\Gamma_-} {\rm d}\hat{n}G_{\hat{n}}\, .
}
The survival probability $P(\hat{n})$ is defined as
\equ{
P(\hat{n})=
\begin{cases}
      \frac{1}{3} &\,\, I_-<I_+, \hat{n}\in \Gamma_-, \\
      1-\frac{2I_-}{3I_+} &\,\, I_-<I_+, \hat{n}\in \Gamma_+, \\
      \frac{1}{3} &\,\, I_->I_+, \hat{n}\in \Gamma_+, \\
      1-\frac{2I_+}{3I_-} &\,\, I_->I_+, \hat{n}\in \Gamma_-, \\
   \end{cases}
}
and the asymptotic angular distributions are given by
\equ{
f^a_{\nu_e}(E,\hat{n})=&P(\hat{n})f_{\nu_e}(E,\hat{n})+[1-P(\hat{n})] f_{\nu_x}(E,\hat{n}),\\
f^a_{\bar{\nu}_e}(E,\hat{n})=&P(\hat{n})f_{\bar{\nu}_e}(E,\hat{n})+[1-P(\hat{n})] f_{\bar{\nu}_x}(E,\hat{n}),\\
f^a_{\nu_x}(E,\hat{n})=&\frac{1}{2}[1-P(\hat{n})]f_{\nu_e}(E,\hat{n})+\frac{1}{2}[1+P(\hat{n})] f_{\nu_x}(E,\hat{n}),\\
f^a_{\bar{\nu}_x}(E,\hat{n})=&\frac{1}{2}[1-P(\hat{n})]f_{\bar{\nu}_e}(E,\hat{n})+\frac{1}{2}[1+P(\hat{n})] f_{\bar{\nu}_x}(E,\hat{n})\,.\\
\label{eq:formula}
}
The following formula is used to estimate the local growth rate of fast flavor {instability} \citep{morinaga2020,nagakura2024}:
\equ{
\sigma = \sqrt{I_+I_-}\, ,
\label{eq:growth}
}
and the change in the neutrino angular distribution per simulation timestep $\Delta t$ is; \equ{
f'_{\nu_\alpha} - f_{\nu_\alpha} = -(1-e^{-\sigma \Delta t})(f_{\nu_\alpha}-f^a_{\nu_\alpha})\, .
\label{control}}
The angular moments used by the M1 transport scheme are calculated from such distributions via numerical integrations in the angular space. {With the relaxation time defined by the inverse growth rate, we don't need to make any assumption concerning the radial range where the prescription of flavor conversion is applied. We apply the scheme to the entire simulation domain and let eq. \ref{control} handle the ``freeze-out" of the flavor conversion at large radii where growth rates are low.}

Our implementation of Box3D in F{\sc{ornax}} conserves the sum of the neutrino/(anti-neutrino) lepton numbers to machine accuracy. Since our M1 transport scheme \citep{skinner2019} evolves only the zeroth and first moments of neutrino angular distributions, an angle-dependent closure relation is needed to reconstruct the full angular distributions. For this purpose we use the Minerbo maximum entropy closure \citep{minerbo1978}. This closure assumes that the neutrino angular distribution can be written in a two-parameter functional form $f(\hat{n})=\exp(a\hat{n}\cdot\hat{f}-b)$, where $a$ and $b$ are parameters derived from the zeroth and first angular moments calculated in F{\sc{ornax}} and $\hat{f}$ is the direction of the neutrino flux. We have in the past explored the dependence of core-collapse simulations without oscillation effects on the second- and third-moment closure relations employed in M1 and found very little variation \citep{wang_burrows2023}. We use Lebedev quadrature on the sphere with 110 points to perform the necessary angular integrations, which integrates exactly all spherical harmonics up to 17th order. 
The result of these integrations for every spatial point, neutrino species, energy-group, and timestep is a slowdown of F{\sc{ornax}} by a factor of $\sim$2.0.

{As is commonly done in sophisticated multi-D CCSN codes to improve the speed (e.g. \citet{ehring_help_or_hinder2023,mori2025_ffc}), F{\sc{ornax}} uses a ``3-species" scheme which use one ``$\mu$-type" neutrino type to represent all heavy neutrino and anti-neutrino species. We note that this scheme assumes} heavy-lepton flavor neutrinos and anti-neutrinos have equal distributions ($f_{{\nu}_x}=f_{\bar{\nu}_x}$). {To update this ``$\mu$-type," we take the average of the last two equations in eq. \ref{eq:formula}}. 
This assumption allows the indirect mixing between $\nu_e$ and $\bar{\nu}_e$ via $\nu_e\leftrightarrow\nu_x=\bar{\nu}_x\leftrightarrow\bar{\nu}_e$ {and violates the ELN-XLN strict conservation. However, we argue that the errors introduced by this assumption are tolerable. In our simulations, there is never a deep crossing or strong conversion, because anytime a crossing is created it is quickly erased by the FFC scheme. Therefore, the XLN should never deviate from zero too much and the error introduced by the assumption in any timestep is actually much smaller than the uncertainty of the angular reconstruction. The question is whether or not this error would accumulate over time, and the way to test this is to follow the total ELN ($N_{\nu_e}-N_{\bar{\nu}_e}$) over a long period of time, because if the error accumulates it will continuously mix $\nu_e$ and $\bar{\nu}_e$. We check the total ELN in 1D in the pre-shock region so that we can ignore the effect of neutrino-matter interactions. We find that total ELN is changed by no more than 10\% (more discussions can be found in section \ref{1D_comp}). Therefore, we suggest this is a good first-order approach. However, to test the accuracy of this method, it would need to be compared with 4-species CCSN simulations in the future. }

\section{Results}  
\label{results}

\subsection{Comparison of One-Dimensional Models: With and Without Fast-Flavor Conversion}
\label{1D_comp}

We have calculated for this study models in both 1D (spherical) and 2D (axisymmetric), with and without fast-flavor conversion \`a la our FFC scheme.  We highlight a few representative progenitors (z9.6, s9.0, {s12.25, s14, s18} and s25) taken from A. Heger (z9.6, private communication) and \citet[][s9.0, {s12.25, s14, s18, s25}]{swbj16,sukhbold2018}. These are a few of the same models employed in our previous 3D studies     \citep[e.g.,][]{burrows_correlations_2024,wang_low_2024}. The goals of this investigation are twofold: First, to determine the alterations in the neutrino fields in time and space due to the FFC. Second, to determine the hydrodynamic consequences of these variations. The latter would ostensibly result from changes in the neutrino heating rates due to changes in energy spectra and species mix and due to changes in the shrinkage rate of the proto-neutron star (PNS) core, with the former effect likely predominant.  We note that ours are the first CCSN calculations to incorporate a reasonable algorithm for the FFC naturally into a supernova code that also automatically addresses the hydrodynamic and neutrino transport feedbacks {without the need for ad hoc free parametrizations.}

Generally, as the left hand side of Figure \ref{fig:heating} demonstrates, there is a delay of $\sim$tens of milliseconds (ms) in the onset of the effect of the FFC on the net neutrino heating rate, which lasts for a finite time  (hundreds of milliseconds). During this phase the net heating rate in the gain region can {be altered by} $\sim$10\%. {Low mass models show enhanced heating rates, while massive ones show less enhanced heating rates and even slightly decreased heating rates.} After that interval, the heating rate behind the shock settles to that without the FFC for the lowest-mass progenitors, but generally lower for the higher-massprogenitors. In 1D, the shock radius, due mostly to the slightly enhanced heating rate behind it, achieves a slightly larger radius (by a few to ten percent), and then recedes in a manner reminiscent of models without the FFC. The right panel of Figure \ref{fig:heating} demonstrates this behavior. 

Interestingly, even after the heating enhancement in the interior subsides, in the region exterior to the stalled shock and at larger distances the FFC can still operate, particularly as the radiation field becomes more forward-peaked; the mix of neutrino species is altered from what it would be without the FFC.  Figure \ref{fig:luminosities} demonstrates the time evolution in 1D of the energy (left) and number (right) luminosities at 10000 kilometers (km) with (solid) and without (dashed) the FFC. Asymptotically, the ``$\nu_{\mu}$" luminosities (blue) are enhanced by a few to $\sim$10\% for the low-ZAMS-mass 9.0 $M_{\odot}$ progenitor (tapering off to no effect after $\sim$500 ms) and by as much as a factor of two for the 18 $M_{\odot}$ model (not tapering off during the simulation). The corresponding effects of the FFC on the $\nu_e$ (red) and $\bar{\nu}_e$ (green) neutrinos are in the opposite direction and of comparable fractional magnitude. They also persist for the more massive progenitor. The differences between the behavior for the 9 and 18 $M_{\odot}$ models can be traced to the larger mass accretion rates for the more massive (higher compactness \citep{2011ApJ...730...70O}) star and its continued accretion in 1D. 

In addition, species equipartition is not achieved in the gain region relevant to the explosion. However, and importantly, as the right-hand side of Figure \ref{fig:luminosities} indicates, despite the operation of the FFC, species equipartition is approached, but not completely achieved at 10,000 km and later times. {We note that ``equipartition" in our three neutrino species case is different from full equipartition. If we had distinguished $\nu_x$ and $\bar{\nu}_x$ neutrinos, flavor equipartition would mean $\nu_e=\nu_x$ and $\bar{\nu}_e=\bar{\nu}_x$. In the three-species case, equipartition means that the two species with lower number densities (in our case $\bar{\nu}_e$ and $\nu_x$) have the same distribution, while the total ELN $N_{\nu_e}-N_{\bar{\nu}_e}$ is conserved. This is the ``equipartition'' described by the phenomenological method in \citet{ehring2023a}, and is very similar to what we see at large radii. {In our scheme, the neutrino mixing is gradually turned off by the growth rate control in equation \ref{control}, and the equipartition is approached but not fully reached.} 

Figure \ref{fig:spectra} compares the neutrino spectra with and without the FFC for the same two progenitors.  When the FFC operates, the $\nu_{\mu}$ neutrino spectra are softer and the $\nu_e$ and $\bar{\nu}_e$ neutrino spectra are harder. The hardening of the spectra of the electron types compensates in the heating rate for the diminution in their number and energy fluxes due to the operation of the FFC (see left panel of Figure \ref{fig:heating}).

Interestingly, Figure \ref{fig:ffc_rate} shows the FFI {growth rate} (eq. \ref{eq:growth}) versus radius at a given time and manifests the roughly power-law behavior at larger radii expected for this quantity \citep{morinaga2020}. This demonstrates in part that our algorithm captures the anticipated behaviors and also shows where the FFC emerges. The major important regions in the context of supernova hydrodynamics where the FFC operates are just behind and just above the shock wave. However, as Figure \ref{fig:ffc_rate} suggests, even when the neutrinos are not coupled to matter the FFC can still occur at larger radii. The FFC does not occur at the base of the gain region, nor in the PNS core.

{Figure \ref{fig:fractions} shows the profile of neutrino energy density fraction of each neutrino type. The region where FFC occurs can be clearly seen and the conversion rate can be estimated. However, we note that the change in such fractions is a joint result of FFC and the feedback from the hydrodynamics and neutrino-matter interactions, and it is difficult to distinguish their individual effects, especially interior to the shock. At early times, most conversion happens at smaller radii (interior to the shock) and equipartition is approached at a few hundred kilometers, while the pre-shock FFI contributes only minor changes. However, as the post-shock FFI region diminishes, the FFC at larger radii becomes more and more important, and equipartition is approached gradually at thousands of kilometers. Note that the difference between the $\nu_e$ and $\bar{\nu}_e$ fraction exterior to the shock can be used to check total ELN conservation. If the $\nu_e\leftrightarrow\nu_x=\bar{\nu}_x\leftrightarrow\bar{\nu}_e$ conversion is very strong, we expect to see a gradually decreasing $\nu_e-\bar{\nu}_e$ when the $\nu_\mu$ fraction changes due to FFC, which is not the case seen here. By checking the neutrino number density profiles, we determine that the total ELN change is within 10\%.}

Model z9.6 is one of the few progenitor models that explodes in 1D \citep{wang_low_2024}. The right panel of Figure \ref{fig:heating} demonstrates that the FFC gives that model a slight boost, causing it to explode a tad earlier and with slightly higher shock speeds. However, the effect in 1D is not large.

\subsection{Comparison of Two-Dimensional Models: With and Without Fast-Flavor Conversion}
\label{2D_comp}    

We now turn to simulations in 2D, with and without fast-flavor conversions and for the same progenitor models discussed in \S\ref{1D_comp}. Without the FFC it has long been known that the neutrino-driven turbulence behind the shock wave that emerges in multi-D is central to the explosion of theoretical core-collapse supernovae \citep{bhf1995,janka1996,janka2012,burrows2013,burrows_vartanyan_nature}. Under this paradigm, most multi-D models now explode effortlessly \citep{burrows_correlations_2024}. However, how might the FFC alter this narrative?  What are the potential effects of flavor changes and neutrino mixing on supernova observables and on the emitted neutrino luminosities themselves? In the context of the FFC and embedding the approach found in \citet{richers_box3D.2024} into our workhorse multi-D supernova code F{\sc{ornax}} \citep{skinner2019}, we are now in a position to address these questions, if only preliminarily.  

Figure \ref{fig:heating-2d} depicts on the left the heating rate in the gain region and on the right the mean shock radius versus time. What we see is that the heating rates can differ appreciably for the higher-mass progenitors, while only modestly for the lower-mass progenitors. At early times for the lower-mass progenitors, the heating rates are marginally higher and the evolution of the shock radius reflects this (see also Figure \ref{fig:heating}).
However, for the higher mass progenitors the heating rates at early times for these 2D models are indeed higher. But, since these models generally explode (if they do) later on, it is the heating rates at later times that are germane and these with the FFC are at times lower, at times higher after $\sim$300 ms after bounce.  One result is that the 25-$M_{\odot}$ model that explodes in 2D and 3D in our previous work doesn't here explode in 2D.  Whether this is the case in 3D and at higher resolution \citep{hiroki_2019}, or whether this is some consequence of chaotic stochasticity when the explodability is close, remains to be determined. Nevertheless, this will be an intriguing topic for future exploration.
 
Figure \ref{fig:luminosities-2d} is the 2D version of Figure \ref{fig:luminosities} and shows quite similar behavior. At late times, equipartition seems to be approached, but at 10000 km is not reached. Though the $\bar{\nu}_e$ and $\nu_{\mu}$ species are close to equipartition, the $\nu_e$ neutrinos are separate. Nevertheless, at ``infinite" radius and if we had distinguished in this study $\nu_{x}$ neutrinos from $\bar{\nu}_{x}$ neutrinos, there remains the distinct possibility that at least partial species equipartition 
($\nu_x=\nu_e$ and $\bar{\nu}_x=\bar{\nu}_e$) might indeed have been achieved at later times (see footnote in \S\ref{1D_comp}). If this were the case, the detection and interpretation of supernova neutrinos would be interestingly altered. A very important aspect of future work on the role of the FFC in supernovae will be the investigation of this stunning possibility.     

Figure \ref{fig:spectra-2d} portrays the same shifts in 2D seen in Figure \ref{fig:spectra} for 1D. The FFC hardens the $\nu_e$ and $\bar{\nu}_e$ neutrinos, while softening the $\nu_{\mu}$ neutrinos. Figure \ref{fig:ffc_region_2d} depicts the regions in 2D where the FFC operates. At early times, the {FFI} is clearly more prominent interior to the shock in the gain region, while at later times it does not disappear, though it is more complexly distributed. It remains to be seen what obtains in 3D and when more precise angular information is available. In the interim, the results we obtain are quite interesting and exciting, however preliminary. 

{Figure \ref{fig:fractions-2d} shows very similar behavior in 2D to that depicted in 1D in Figure \ref{fig:fractions}. Since the shock radii are larger in 2D, the FFC also operates at larger radii in 2D. As the post-shock FFI region diminishes, the FFC at larger radii becomes more and more important. This trend is not influenced by the explosion. The differences between $\nu_e$ and $\bar{\nu}_e$ fractions change more than in 1D because of the averages. The total ELN is conserved as well as in 1D along each radial line along which neutrinos propagate.}
       
\section{Conclusions}
\label{conclusions}

In this paper, we have explored the effects of fast-flavor conversion in state-of-the-art 1D and 2D core-collapse supernova simulations incorporating the Box3D algorithm of \citet{richers_box3D.2024} into the F{\sc{ornax}} code. We find that after a few tens of milliseconds after bounce the FFC emerges in strength just interior to and just exterior to the stalled shock wave. It does not obtain in the PNS core nor near the average neutrinosphere radii. Interior to the shock, this results in a temporary increase in the net neutrino heating rate, due mostly to a hardening of the $\nu_e$ and $\bar{\nu}_e$ neutrino spectra despite the decrease in their corresponding neutrino number fluxes. In 1D, the hydrodynamic effects of this modest enhancement are not large, with temporary increases in the stalled shock radius by of order ten to twenty kilometers that abate within a few hundred milliseconds. However, for the one model (z9.6) that explodes in 1D without the FFC the corresponding model with FFC explodes slightly earlier and slightly more energetically.

Importantly, though any enhancement in the neutrino heating rate itself is confined to the quoted inner regions and earlier times, the FFC itself continues to operate at larger radii and at later times, though with diminishing effect (see Figure \ref{fig:ffc_rate}). The net result is a shift upward in the asymptotic ``$\nu_{\mu}$" energy and number luminosities and a corresponding shift downward in the same quantities for both the $\nu_e$ and $\bar{\nu}_e$ neutrinos. In our simulations, this mixing brings the number luminosities closer to, but not to, equipartition at larger radii and later times. The magnitude of the FFC-caused shift in these quantities at a given radius is smaller for the lower-mass/lower-compactness progenitors and larger for the higher-mass/higher-compactness progenitors. However, we speculate that had we distinguished $\nu_{x}$ and $\bar{\nu}_{x}$ neutrinos and extended our simulation space to much larger radii at least partial species equipartition ($\nu_x=\nu_e$ and $\bar{\nu}_x=\bar{\nu}_e$) might have emerged. If this is true, it could interestingly influence the detection of supernova neutrinos in underground and under-ice facilities.  

In 2D, the hydrodynamic effect of the FFC is a bit more noticeable.  For the lower-mass progenitors, turning on the FFC results in slightly earlier explosions than for models that exploded without the fast-flavor effect. We note that such models as the 12.25- and 14-$M_{\odot}$, that did not explode in previous calculations in 2D or 3D \citep{burrows_channels_2024}, still don't explode when the FFC is enabled. Whether the same will be true in 3D remains to be seen. Curiously, for higher-mass progenitors (such as the s25 model) in 2D the FFC seems to inhibit explosion. Whether this remains the case in 3D is not yet clear. It will be important to ascertain the effects of the FFC on the raft of supernova observables, including the final explosion energies and nucleosynthetic yields. The latter depend upon the asymptotic electron fraction ($Y_e$), which is determined by the countervaling effects of $\nu_e$ and $\bar{\nu}_e$ absorption altered slightly by the FFC-induced shifts in 
their respective emission characteristics.

{It's very interesting to compare our self-consistent simulation results to models using a phenomenological method \citep{ehring2023a,ehring_help_or_hinder2023,mori2025_ffc}. This phenomenological method assumes that flavor equipartition is instantaneously achieved below a density threshold $\rho_c$, which is treated as a free parameter. Although the asymptotic neutrino signals are similar (approaching equipartition at large radii), our models show significantly weaker FFC effects on CCSN hydrodynamics. There are two main reasons: First, the size of FFC region is overestimated by the phenomenological methods. In our simulations, the FFC region interior to the shock where neutrinos are coupled to hydrodynamics diminishes quickly, while a fixed density threshold can not correctly capture the change in the FFC region over time and will overstate its size at later times. In addition, there is a delay after bounce before the FFC occurs, which is also not captured by the density threshold method. Second, the equipartition state is not instantaneously achieved. From Figures \ref{fig:fractions} and \ref{fig:fractions-2d}, we see that most often flavor equipartition is gradually approached exterior to the shock, where the neutrino-matter coupling is weak. The phenomenological method assumes flavor equipartition is achieved instantaneously at some radius interior to the shock where neutrinos are still coupled with matter, which could easily overestimate the impact on CCSN hydrodynamics.
}

As suggested by \citet{ehring_help_or_hinder2023}, the residual neutron star mass of the lowest-mass progenitors may be shaved a bit ($\sim$0.02 $M_{\odot}$) relative to that without the FFC.
This might help reconcile the apparent gap between the theoretically predicted lowest mass neutron star ($\sim$1.2-1.25 $M_{\odot}$, \citep{radice2017b}) and the lowest measured mass for a neutron star (1.174 $M_{\odot}$, \citep{Martinez_2015}).  {Although we find that the FFC effects on CCSN hydrodynamics are weaker than expected in \citet{ehring2023a,ehring_help_or_hinder2023,mori2025_ffc}, models with the lowest progenitor masses might still be interestingly influenced and further studies are certainly called for.}

The calculations presented in this paper are the first of their kind that couple a classical {time-dependent} radiation/hydrodynamics supernova code with a credible algorithm for incorporating fast-flavor conversion between neutrino species to determine the consequences for both supernova hydrodynamics and the emergent neutrino signal. Our results are interesting and provocative, but they are only preliminary {and come with caveats}. The Box-like method in \citet{zaizen2023} is based on QKE simulations with periodic boundary conditions, and the dependence of our results on the boundary conditions needs to be explored \citep{zaizen2023b,cornelius2024}. \tianshu{In addition, the Box-like method makes assumptions that there is only one crossing in angular space and that the survival probability function is piecewise constant. Such assumptions need to be further tested and improved.} The assumption that $\nu_x=\bar{\nu}_x$ introduced by the 3-species scheme slightly breaks ELN-XLN conservation\tianshu{, and a 4-species scheme, though more expensive, is certainly called for in the future}. \tianshu{Furthermore, the angular distribution reconstruction based on the Minerbo closure (or any other closure relations) hasn't been well-tested and the uncertainties introduced by such reconstructions remain.} There is also an inconsistency between the Minerbo closure reconstruction of neutrino angular distributions and the angular distribution given by the Box3D method \citep{richers_box3D.2024}, which needs to be studied. {In addition,} a new series of 3D simulations incorporating the FFC for a range of progenitors will be important next steps. Moreover, collisional flavor transformation \citep{johns_collisional2023}
and spectral swaps \citep{Duan2007}
have yet to be similarly vetted. {Furthermore, slow flavor conversion is another process to consider, though its effects on CCSN hydrodynamics might be quite minor \citep{stapleford_kneller2020}.} In any case, however imperfect, the insights we have garnered here should now serve to ignite the next generation of studies on this exciting topic at the intersection of supernova theory and neutrino physics.

\section*{Data Availability}  

\adam{The data presented in this paper can be made available upon reasonable request to the corresponding author.}

\section*{Acknowledgments}

We thank David Vartanyan for our long-term productive collaboration and Lucas Johns and Hiroki Nagakura for their many insights into the fast-flavor conversion phenomenon. TW acknowledges support by the U.~S.\ Department of Energy under grant DE-SC0004658, support by the Gordon and Betty Moore Foundation through Grant GBMF5076 and Simons Foundation grant (622817DK). AB acknowledges former support from the U.~S.\ Department of Energy Office of Science and the Office of Advanced Scientific Computing Research via the Scientific Discovery through Advanced Computing (SciDAC4) program and Grant DE-SC0018297 (subaward 00009650) and former support from the U.~S.\ National Science Foundation (NSF) under Grant AST-1714267. We are happy to acknowledge access to the Frontera cluster (under awards AST20020 and AST21003). This research is part of the Frontera computing project at the Texas Advanced Computing Center \citep{Stanzione2020}. Frontera is made possible by NSF award OAC-1818253. Additionally, a generous award of computer time was provided by the INCITE program, enabling this research to use resources of the Argonne Leadership Computing Facility, a DOE Office of Science User Facility supported under Contract DE-AC02-06CH11357. Finally, the authors acknowledge computational resources provided by the high-performance computer center at Princeton University, which is jointly supported by the Princeton Institute for Computational Science and Engineering (PICSciE) and the Princeton University Office of Information Technology, and our continuing allocation at the National Energy Research Scientific Computing Center (NERSC), which is supported by the Office of Science of the U.~S.\ Department of Energy under contract DE-AC03-76SF00098.

\newpage
\begin{figure*}
    \centering
    \includegraphics[width=0.48\textwidth]{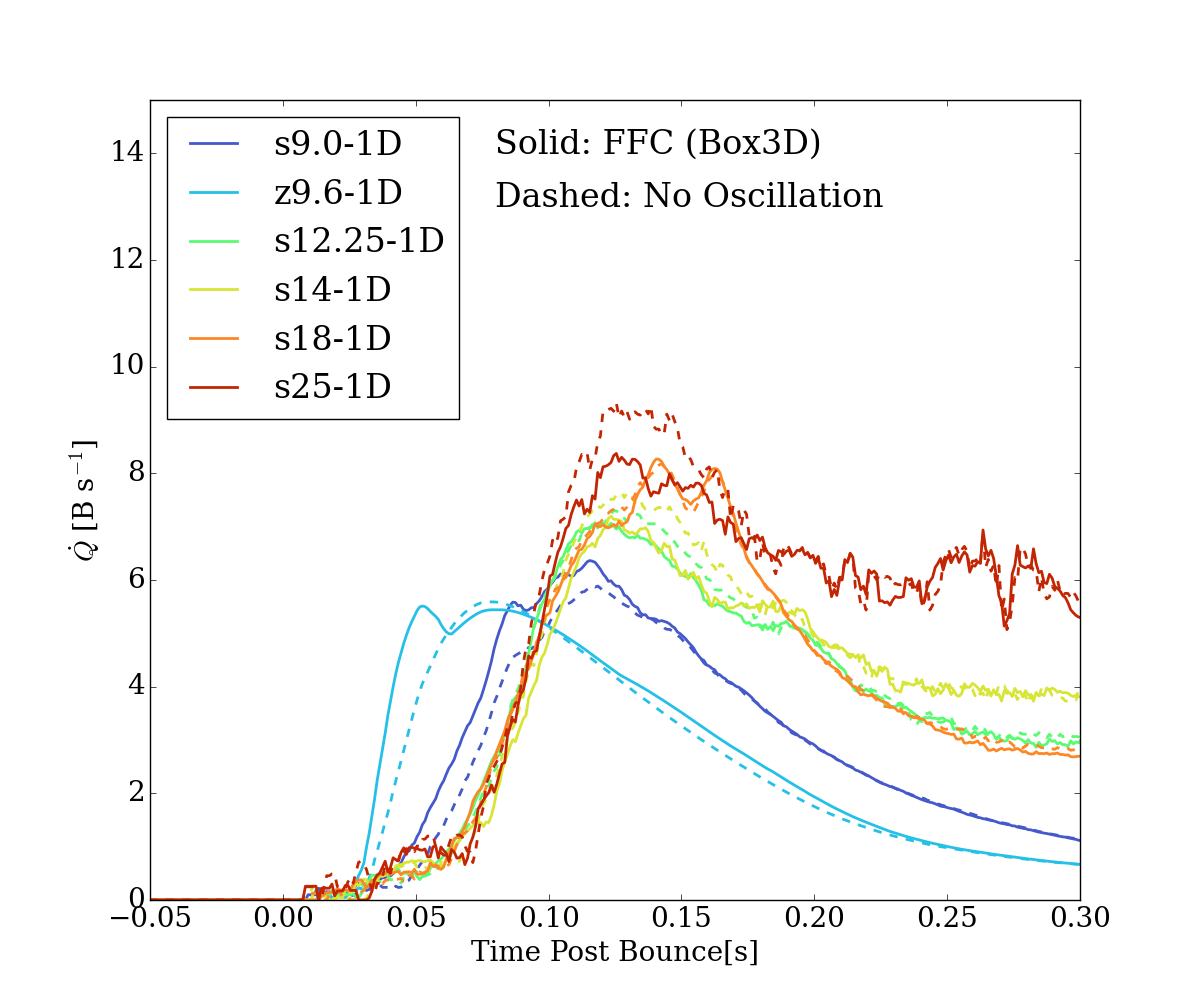}
    \includegraphics[width=0.48\textwidth]{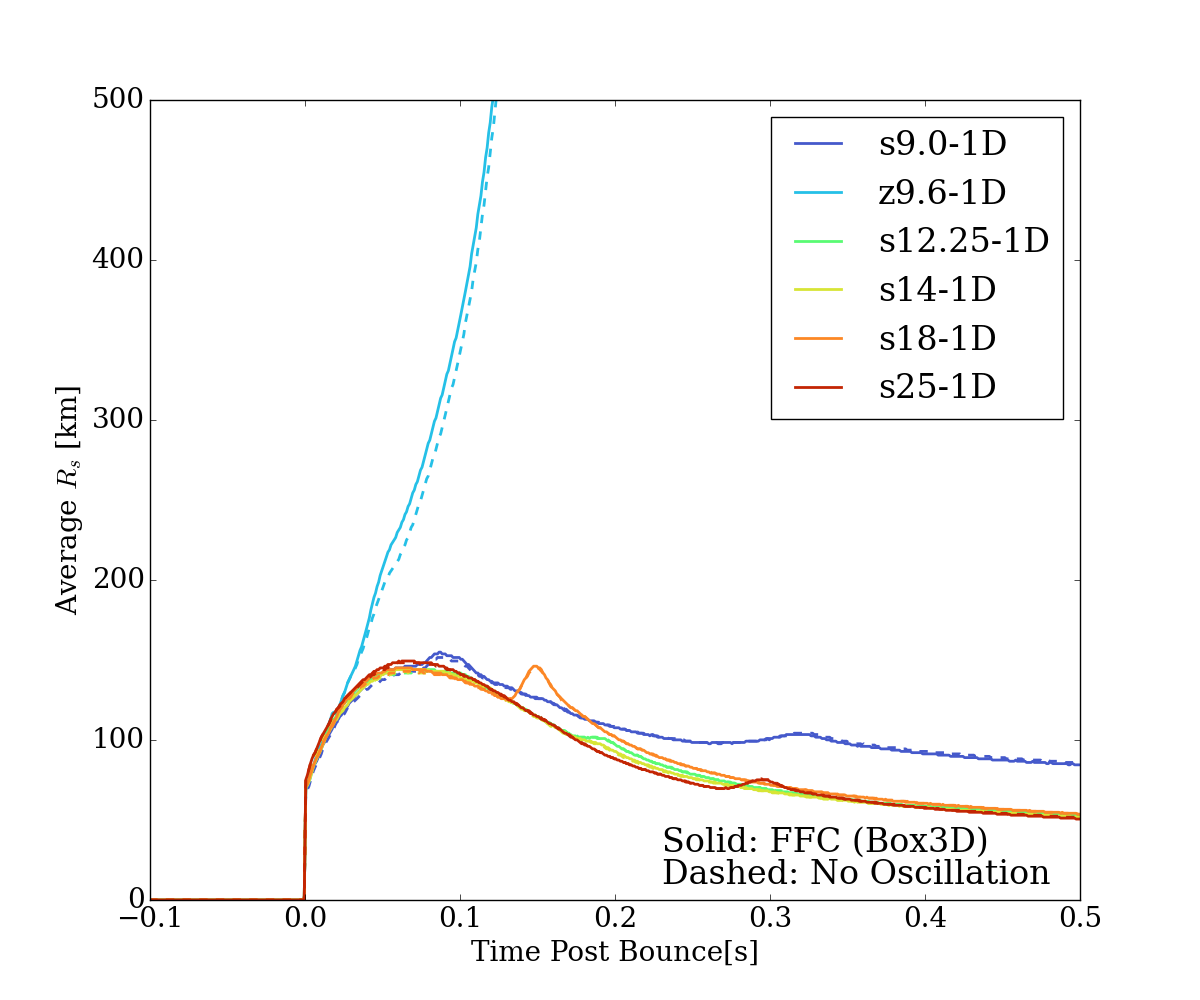}
    \caption{{\bf Left:} The net heating rate (in units of $10^{51}$ erg s$^{-1}$) in the gain region behind the shock for 1D simulations using F{\sc{ornax}} of the z9.6, s9.0, s12.25, s14, s18, and s25 progenitor models versus time (in milliseconds) after bounce.  The dashed curves are without the Box3D implementation of fast-flavor conversion and the solid curves include it.  {\bf Right:} The shock radius versus time after bounce for the models depicted on the left panel of this figure set. In 1D, the inclusion of FFC does not result in a qualitative difference in hydrodynamics, but for the z9.6 model, which does explode in 1D without the FFC, the explosion occurs a bit earlier and a bit more vigorously when FFC effects are included. See text for a discussion of both these panels.}
    \label{fig:heating}
\end{figure*}

\begin{figure*}
    \centering
    \includegraphics[width=0.48\textwidth]{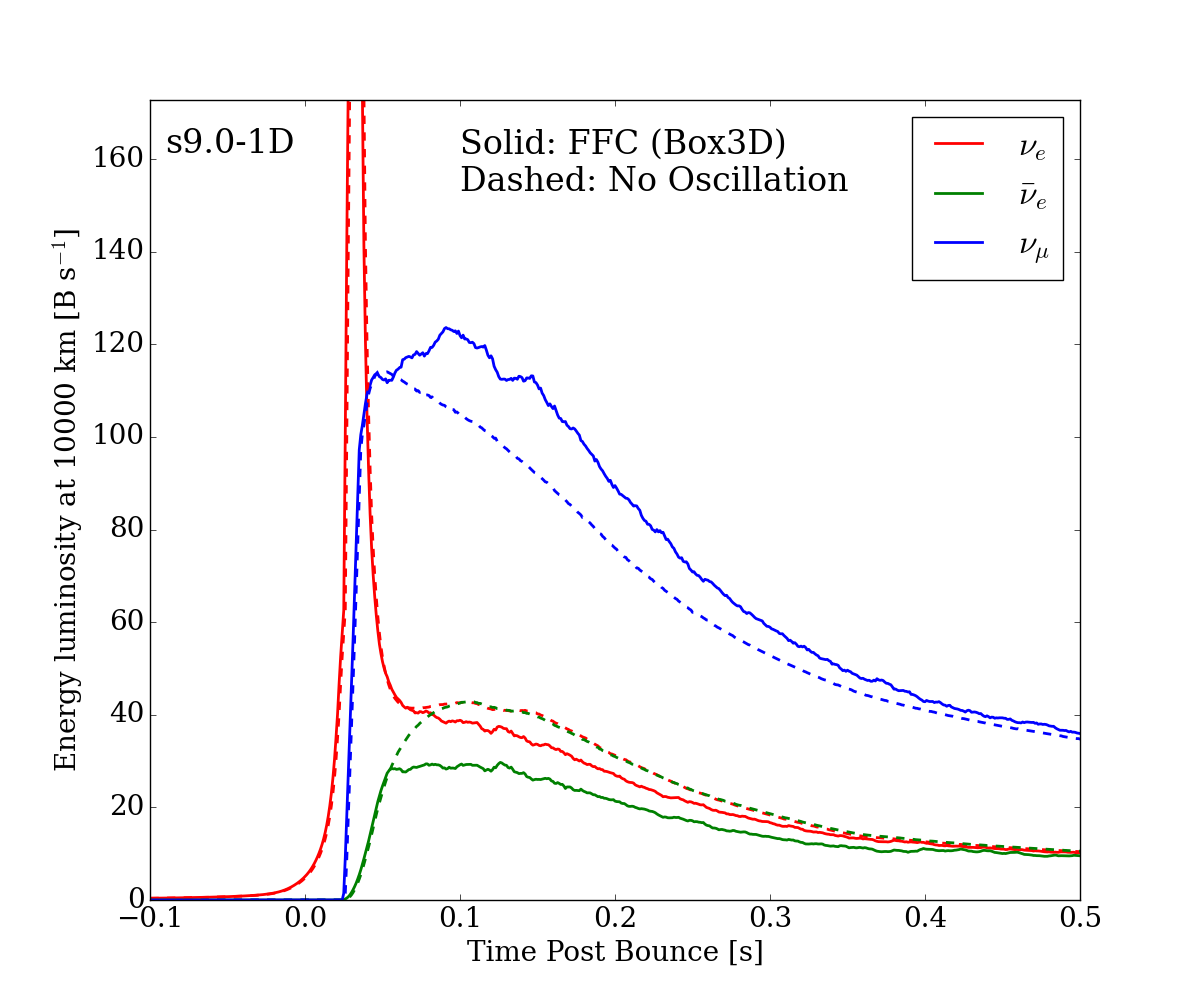}
    \includegraphics[width=0.48\textwidth]{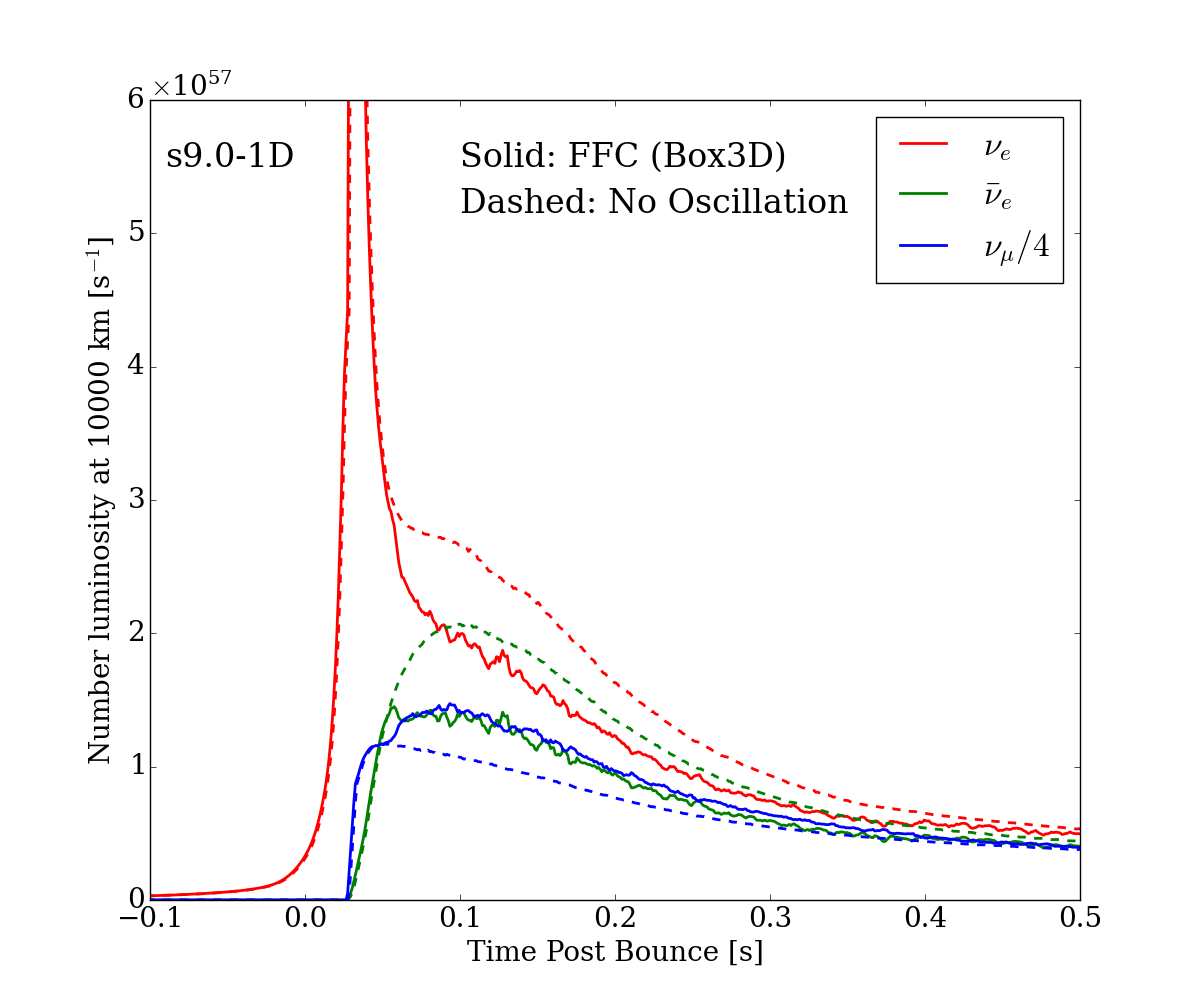}    
    \includegraphics[width=0.48\textwidth]{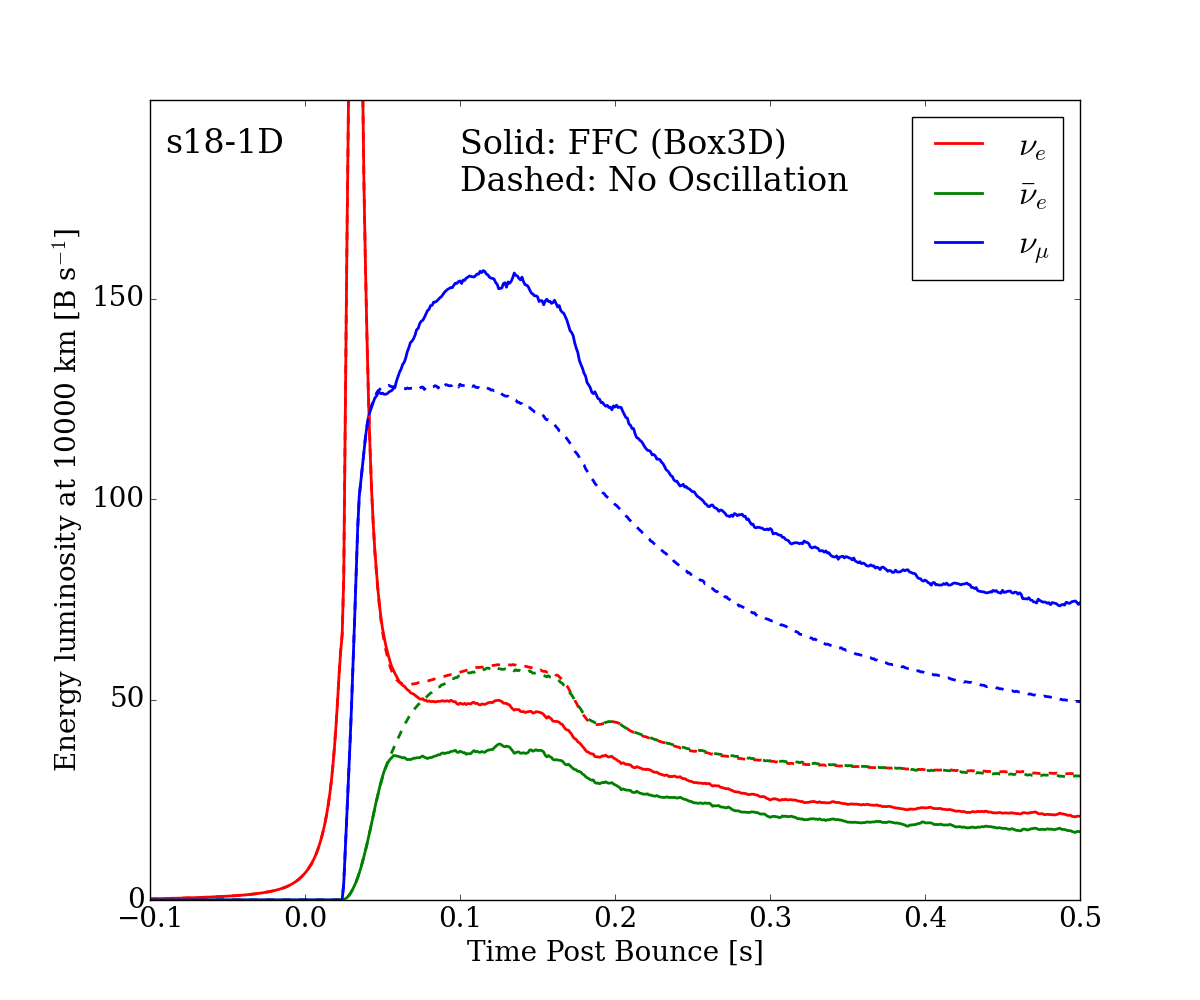}
    \includegraphics[width=0.48\textwidth]{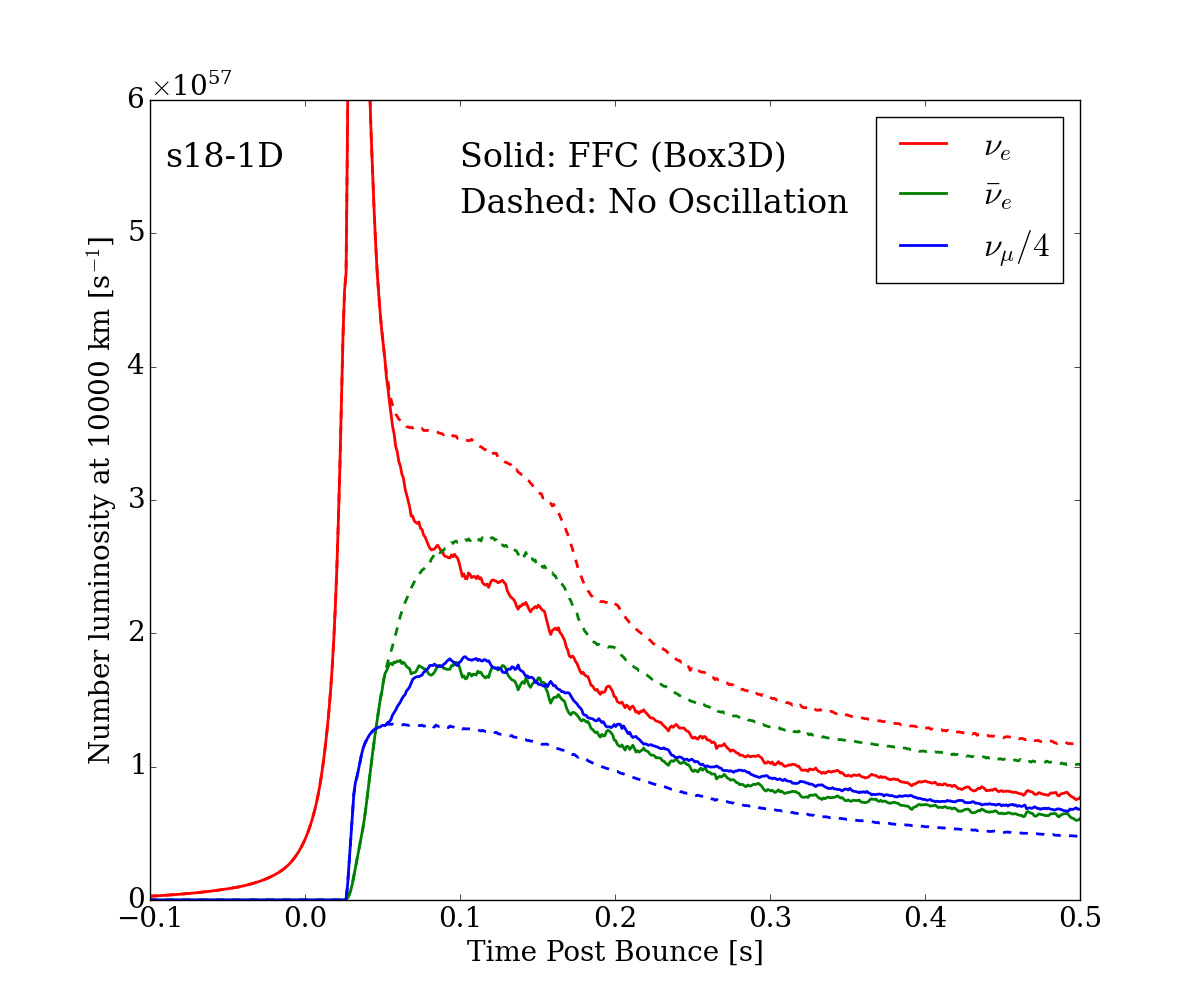}
    \caption{{\bf Left:} The evolution of the neutrino energy luminosities (in units of Bethes ($10^{51}$ erg) s$^{-1}$ at 10000 km) versus time after bounce (in seconds) for the 9 (top) and 18 (bottom) $M_{\odot}$ progenitor 1D simulations of Figure \ref{fig:heating}, with (solid) and without (dashed) the inclusion of the FFC. {Note that $\nu_\mu$ is the combination of $\nu_\mu$, $\bar{\nu}_\mu$, $\nu_\tau$, and $\bar{\nu}_\tau$.} After a slight delay, fast-flavor conversion boosts the $\nu_{\mu}$ luminosities by as much as 20\% for the 9.0 $M_{\odot}$ (and then tapers off) and by $\sim$20\% for the 18 $M_{\odot}$ model (maintaining steady during the simulation). Correspondingly, the $\nu_e$ and $\nu_{\mu}$ luminosities are suppressed, conserving total neutrino number. {\bf Right:} The corresponding number luminosities (in units of $10^{57}$ neutrinos per second) versus time after bounce. {For the number luminosity comparison, we show $``\nu_\mu"/4$.} While the FFC does redistribute the neutrino species mix, equipartition is not fully achieved in these simulations on our finite grid. See text for a discussion.}
    \label{fig:luminosities}
\end{figure*}

\begin{figure*}
    \centering
    \includegraphics[width=0.48\textwidth]{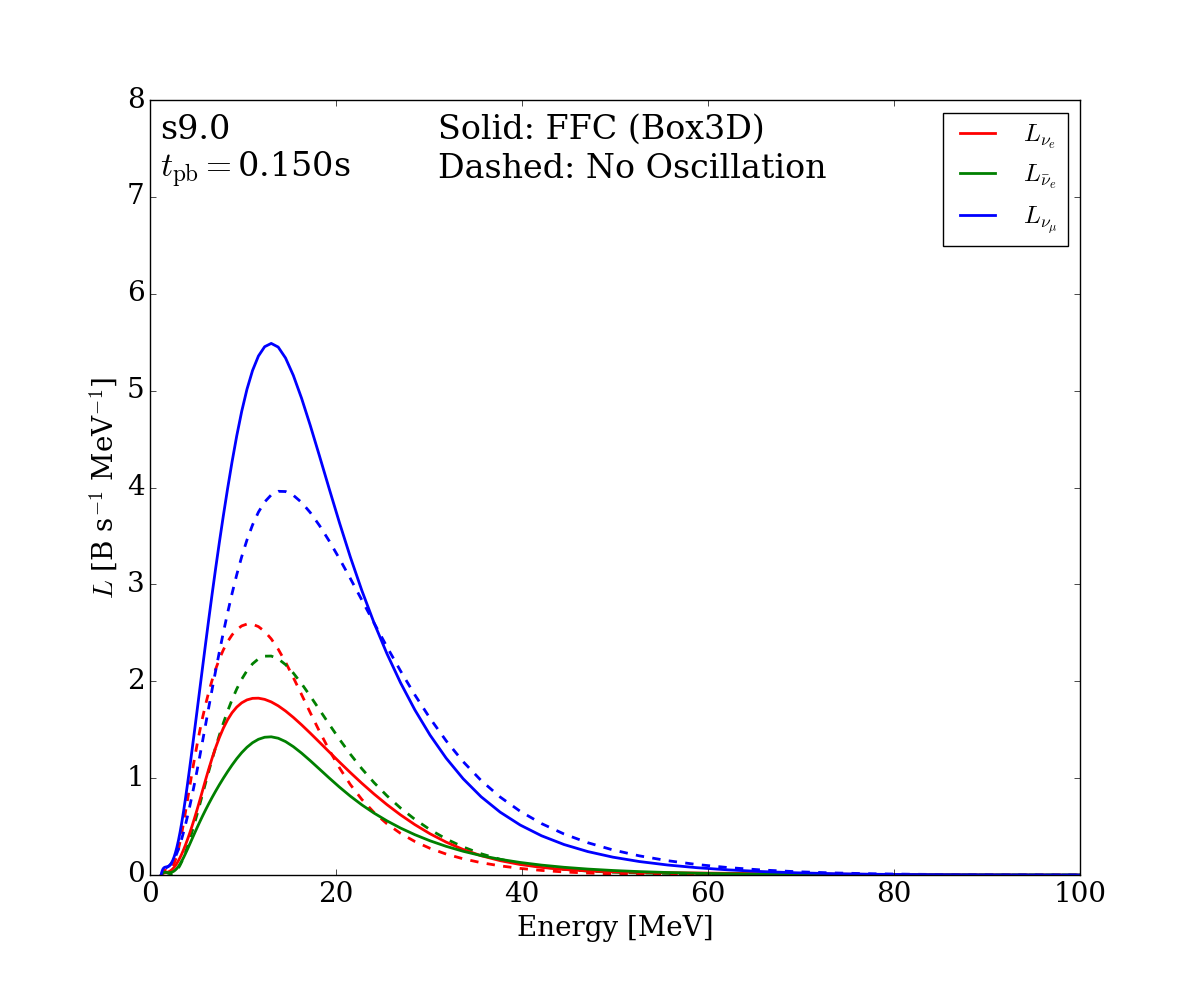}
    \includegraphics[width=0.48\textwidth]{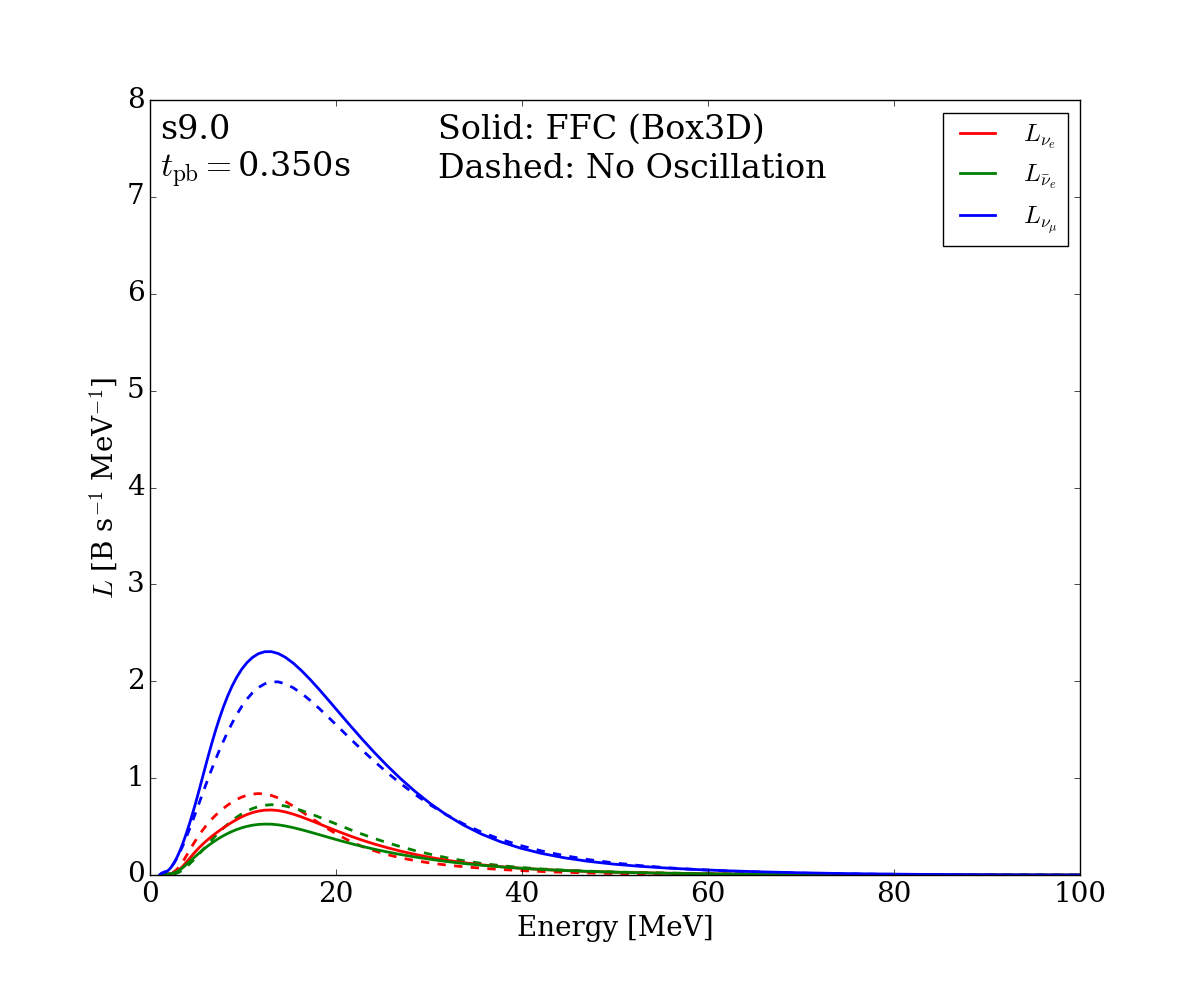}
    \includegraphics[width=0.48\textwidth]{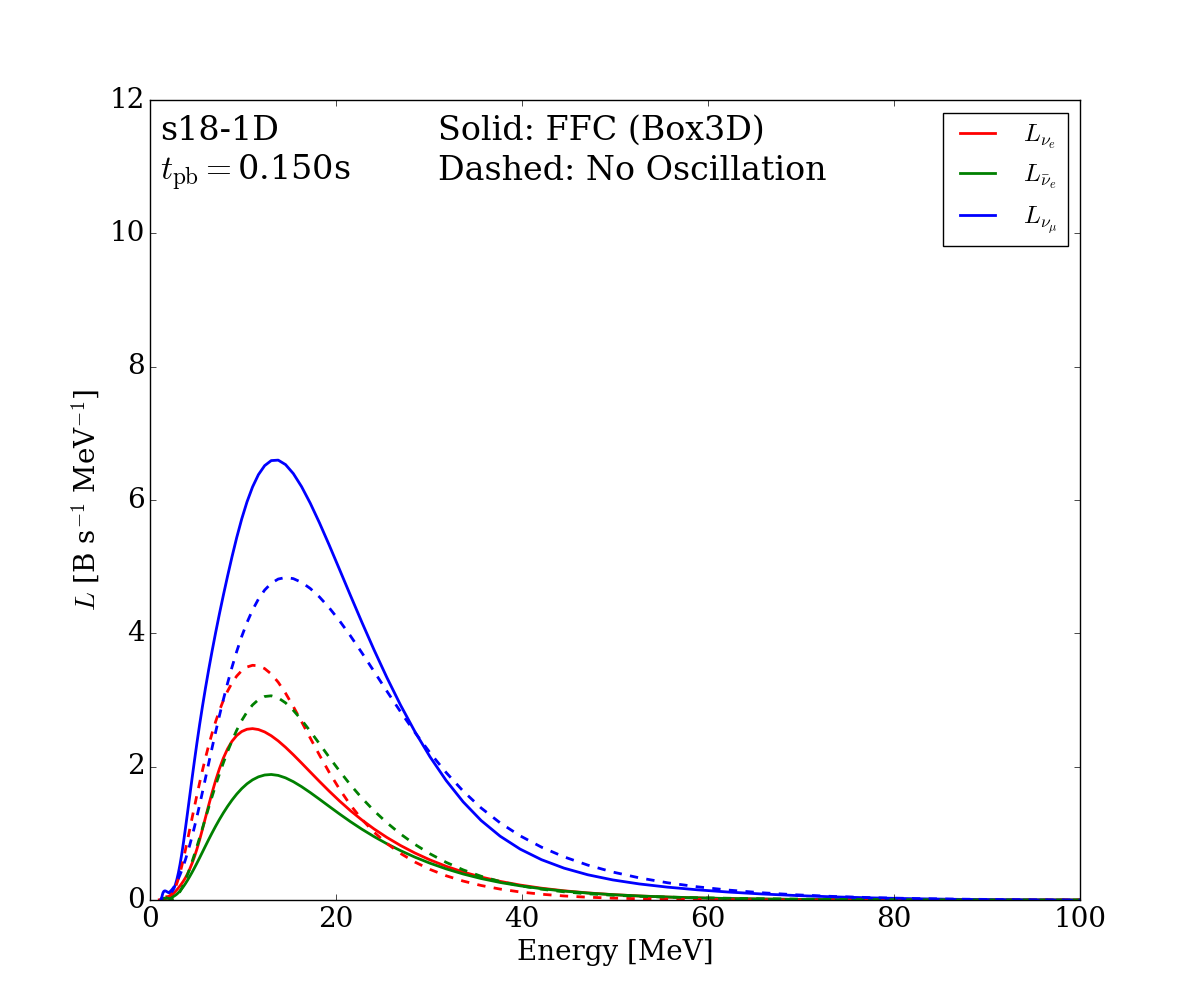}
    \includegraphics[width=0.48\textwidth]{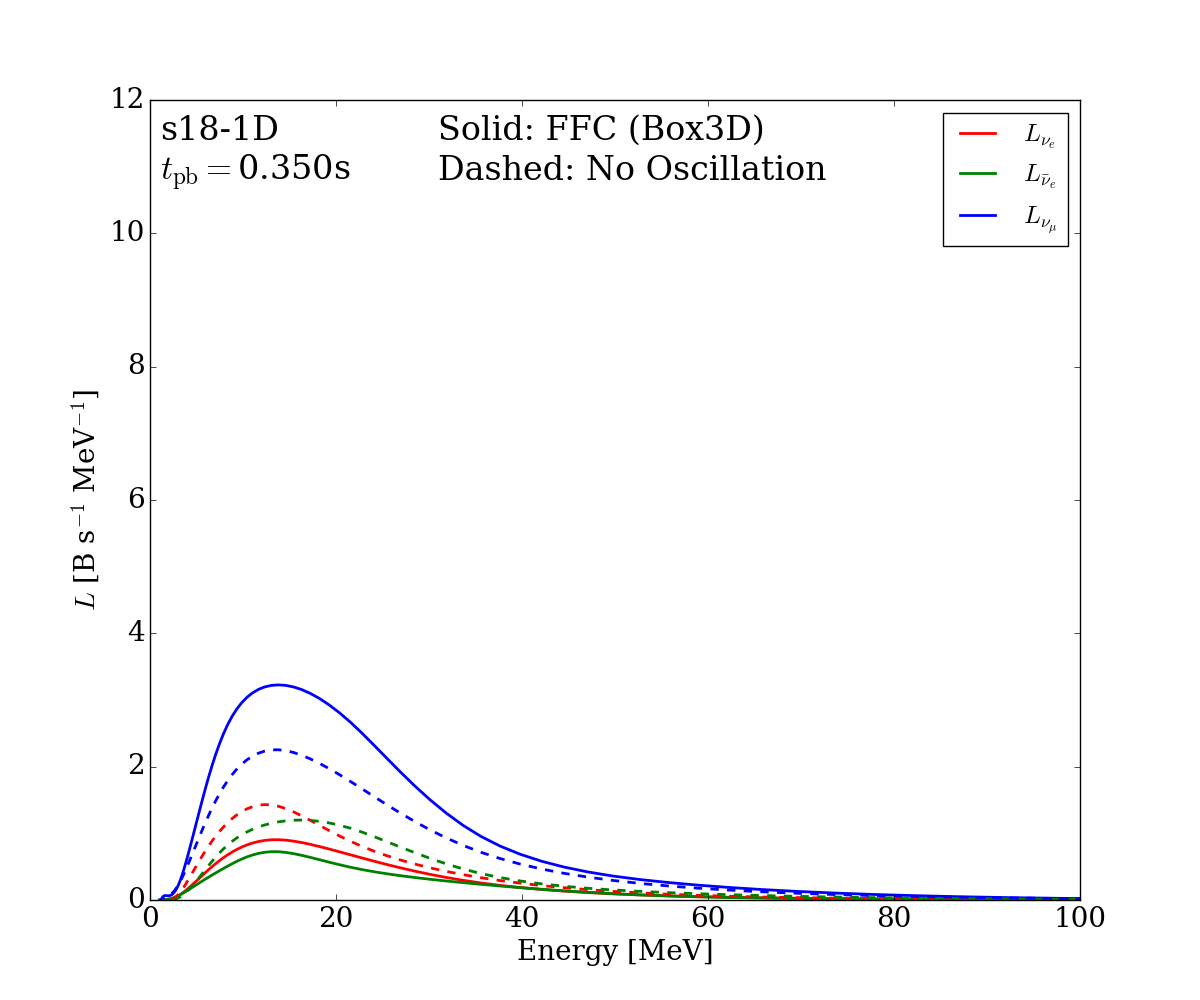}
    \caption{{The emergent spectra (splined) of the neutrinos, with (solid) and without (dashed) the FFC, for 1D simulations of the progenitors highlighted in Figure \ref{fig:luminosities} and for two times after bounce. {Note that $\nu_\mu$ is the combination of $\nu_\mu$, $\bar{\nu}_\mu$, $\nu_\tau$, and $\bar{\nu}_\tau$.} When the FFC is operative, the $\nu_{\mu}$ neutrino spectrum is softened, while both the $\nu_e$ and $\bar{\nu}_e$ neutrino spectra harden slightly.}} 
    \label{fig:spectra}
\end{figure*}

\begin{figure*}
    \centering
    \includegraphics[width=0.48\textwidth]{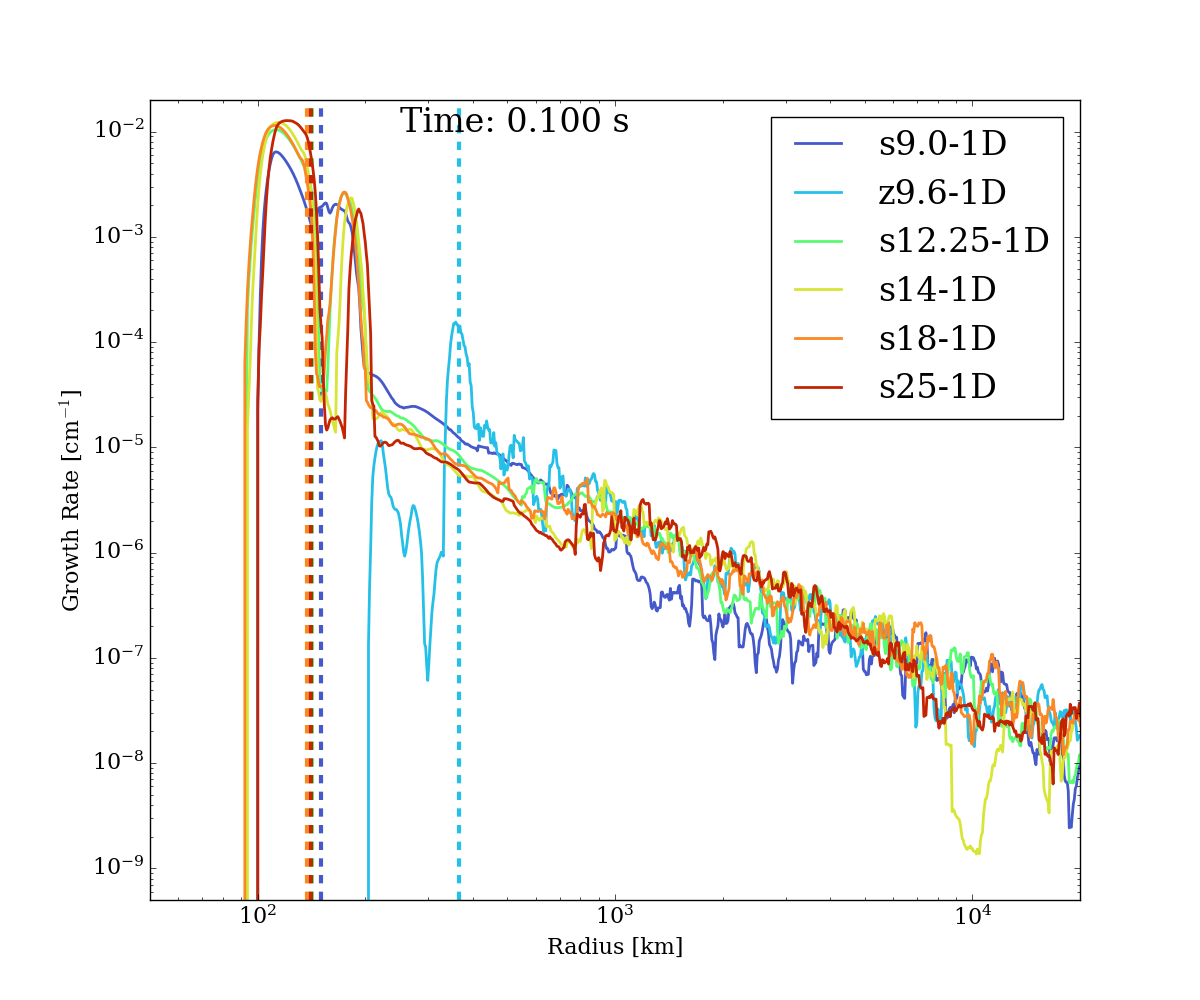}
    \includegraphics[width=0.48\textwidth]{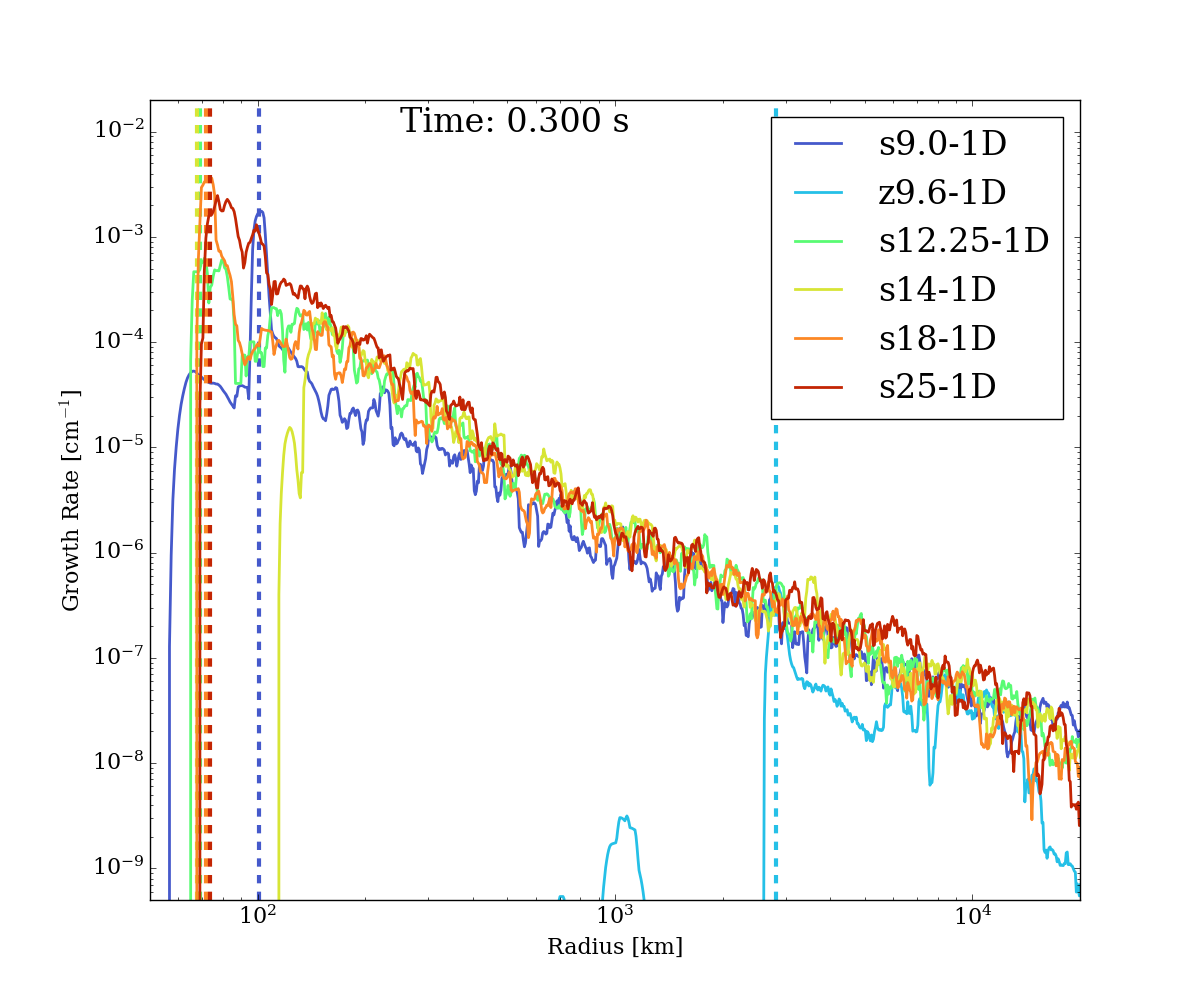}
    \caption{{The local {FFI growth} rate (using eq. \ref{eq:growth}, in units of cm$^{-1}$) versus radius for two representative times (100 and 300 milliseconds) after bounce for representative 1D models when the FFC is operating. We have smoothed these curves with boxcar averages in radius (over 10 km) and in time (over 10 ms).  The region just interior to and just exterior to the shock wave (vertical dashed line) is where the {growth} rate is largest. There is no {FFI} interior to the outer gain region, either in the proto-neutron star or near the average neutrinospheres. However, at large radii, though there is effectively no neutrino matter heating, the {FFI} continues, due in no small measure to the progressively more forwardly-peaked angular distribution of the radiation. As this figure demonstrates, the {FFI growth} rate at large radii follows a power-law relation. The power-law index is close to $-(1+\beta)$ expected in \citet{morinaga2020}, where $\beta$ is the power index of the density profile exterior to the shock ($\rho\propto r^{-\beta}$).}}
    \label{fig:ffc_rate}
\end{figure*}

\begin{figure*}
    \centering
    \includegraphics[width=0.48\textwidth]{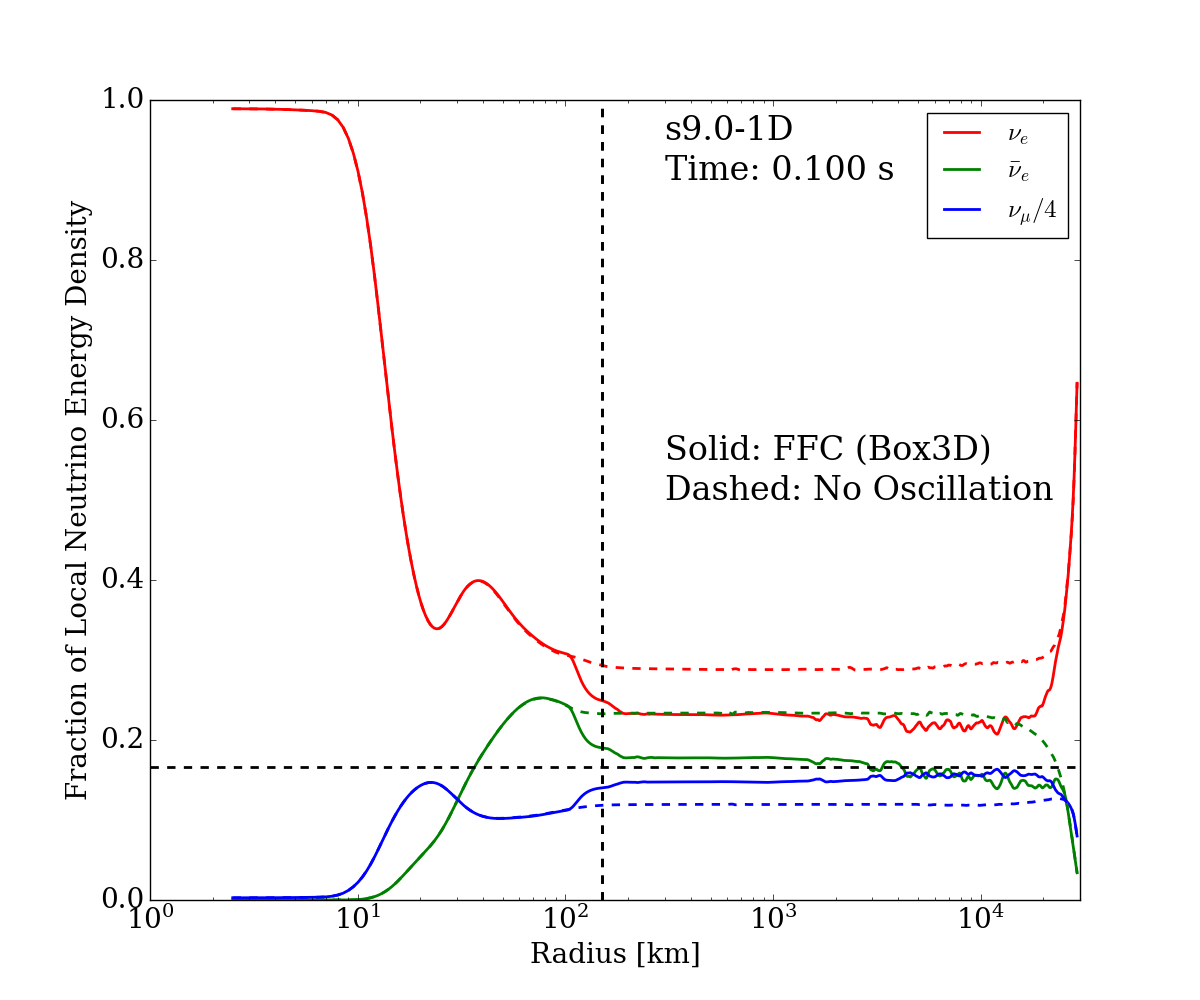}
    \includegraphics[width=0.48\textwidth]{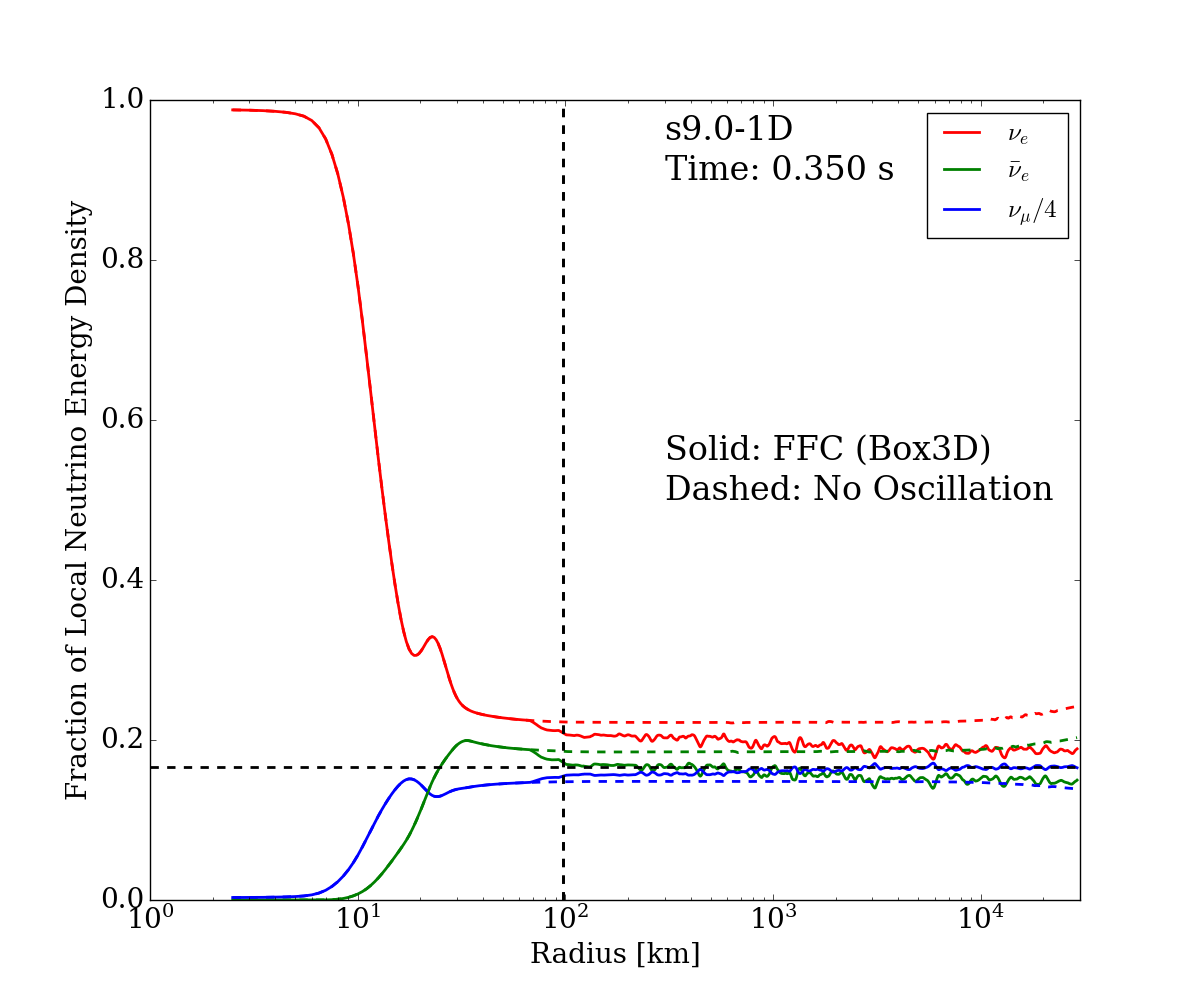}    
    \includegraphics[width=0.48\textwidth]{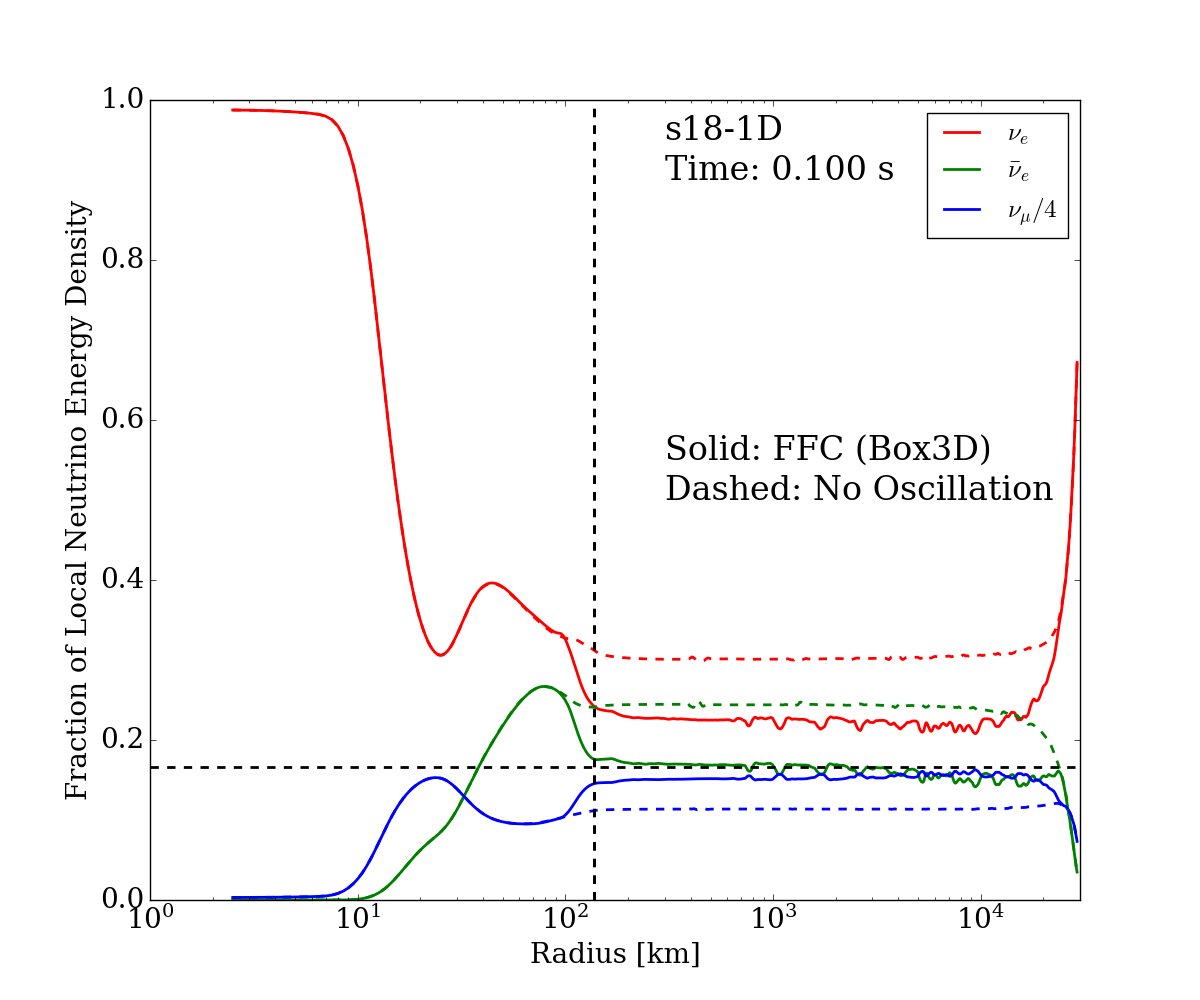}
    \includegraphics[width=0.48\textwidth]{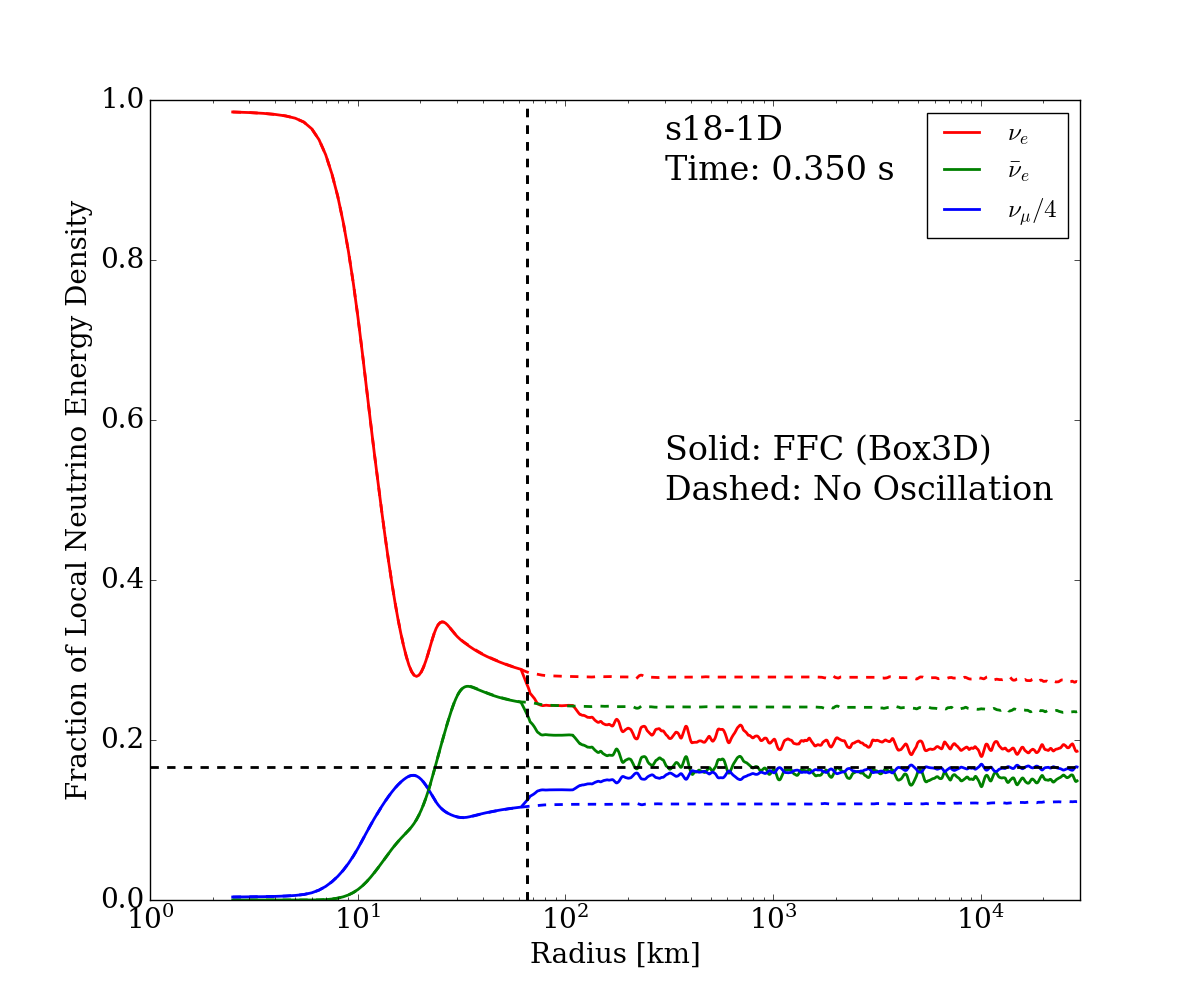}
    \caption{{The neutrino energy density fraction profiles for the 9 (top) and 18 (bottom) $M_{\odot}$ progenitor 1D simulations at two times after bounce, with (solid) and without (dashed) the inclusion of the FFC. Note that $\nu_\mu$ is the combination of $\nu_\mu$, $\bar{\nu}_\mu$, $\nu_\tau$, and $\bar{\nu}_\tau$. We have smoothed these curves with boxcar averages in radius (over 10 km). The vertical dashed line shows the shock radius, while the horizontal dashed line at $1/6\approx0.167$ marks the equipartition fraction. The region where FFC occurs can be clearly seen. At early times, most conversion happens at smaller radii (interior to the shock) and equipartition is approached at a few hundred kilometers, while the pre-shock FFI affects only minor conversions. However, as the post-shock FFI region diminishes, the FFC at larger radii becomes more and more important, and equipartition is approached gradually at thousands of kilometers.}}
    \label{fig:fractions}
\end{figure*}

\begin{figure*}
    \centering
    \includegraphics[width=0.48\textwidth]{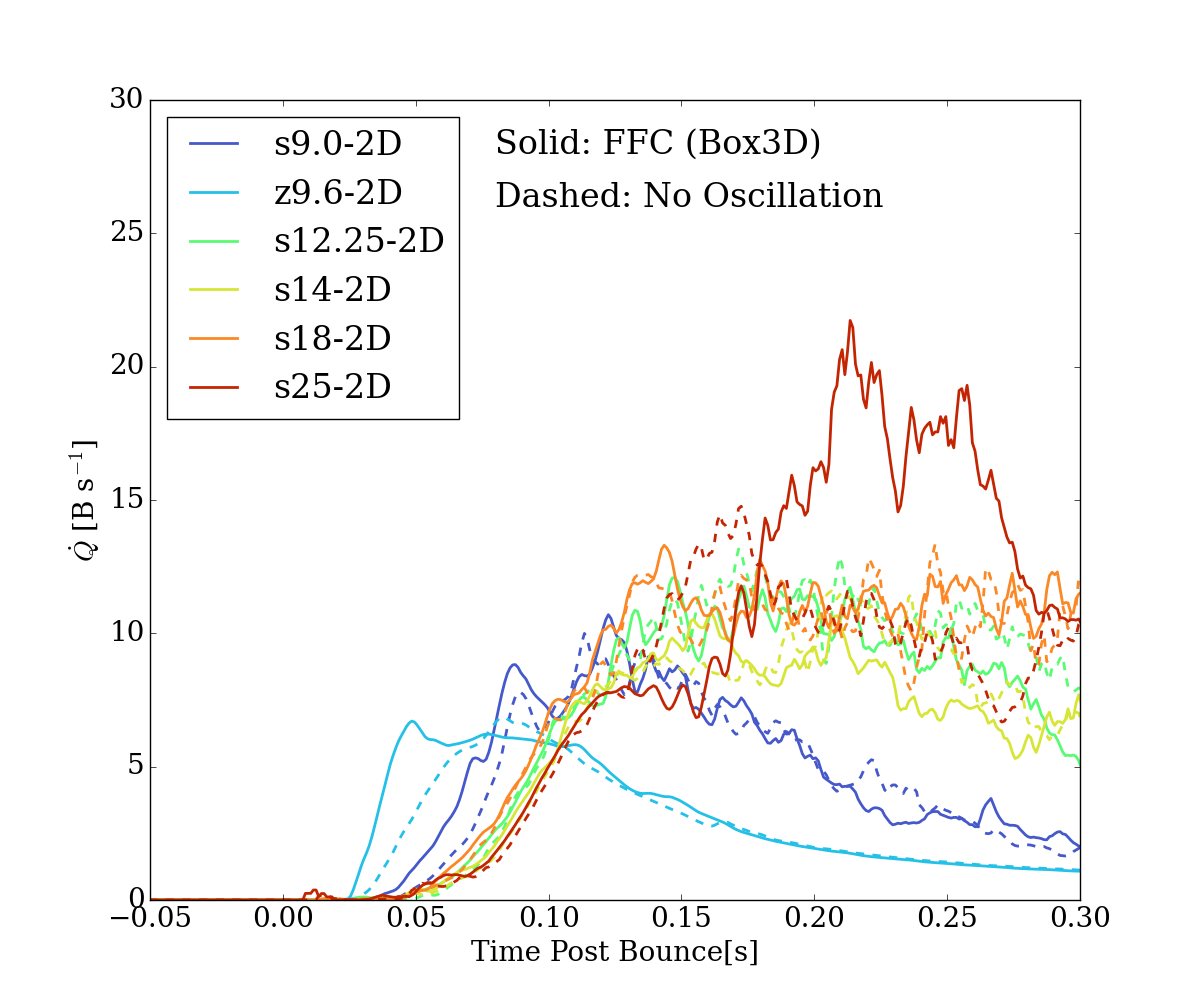}
    \includegraphics[width=0.48\textwidth]{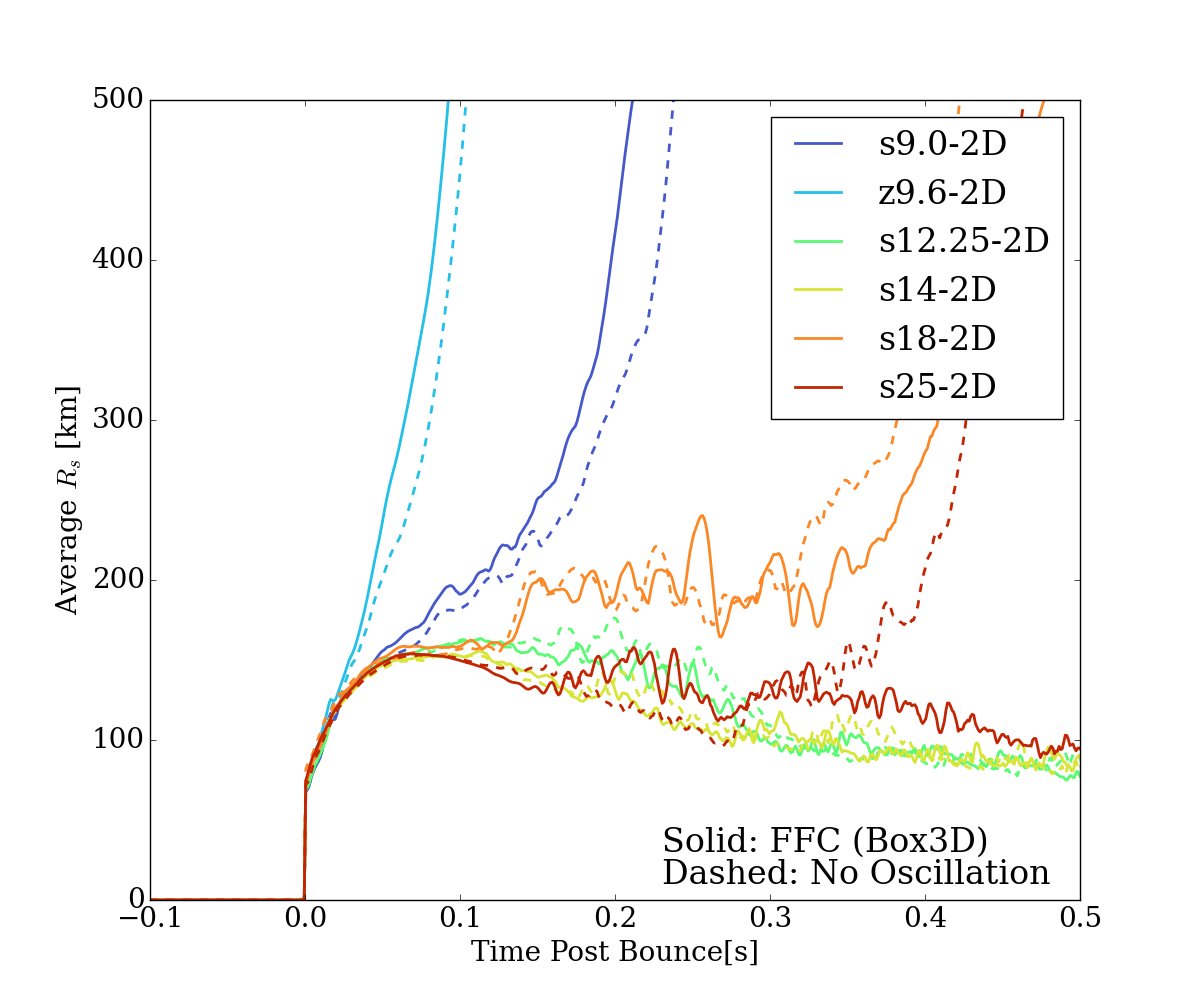}
    \caption{{\bf Left:} The net heating rate (in units of $10^{51}$ erg s$^{-1}$) in the gain region behind the shock for 2D simulations of the z9.6, s9.0, s12.25, s14, s18, and s25 progenitor models versus time (in milliseconds) after bounce.  The dashed curves are without the Box3D implementation of fast-flavor conversion and the solid curves include it.  {\bf Right:} The shock radius versus time after bounce for the models depicted on the left panel of this figure set. The low-mass models explode more promptly with FFC, while the more massive models are less sensitive to it. The s25 model fails to explode if FFC is included. This shows the possibility that FFC may hinder the explosion in more massive models.}
    \label{fig:heating-2d}
\end{figure*}

\begin{figure*}
    \centering
    \includegraphics[width=0.48\textwidth]{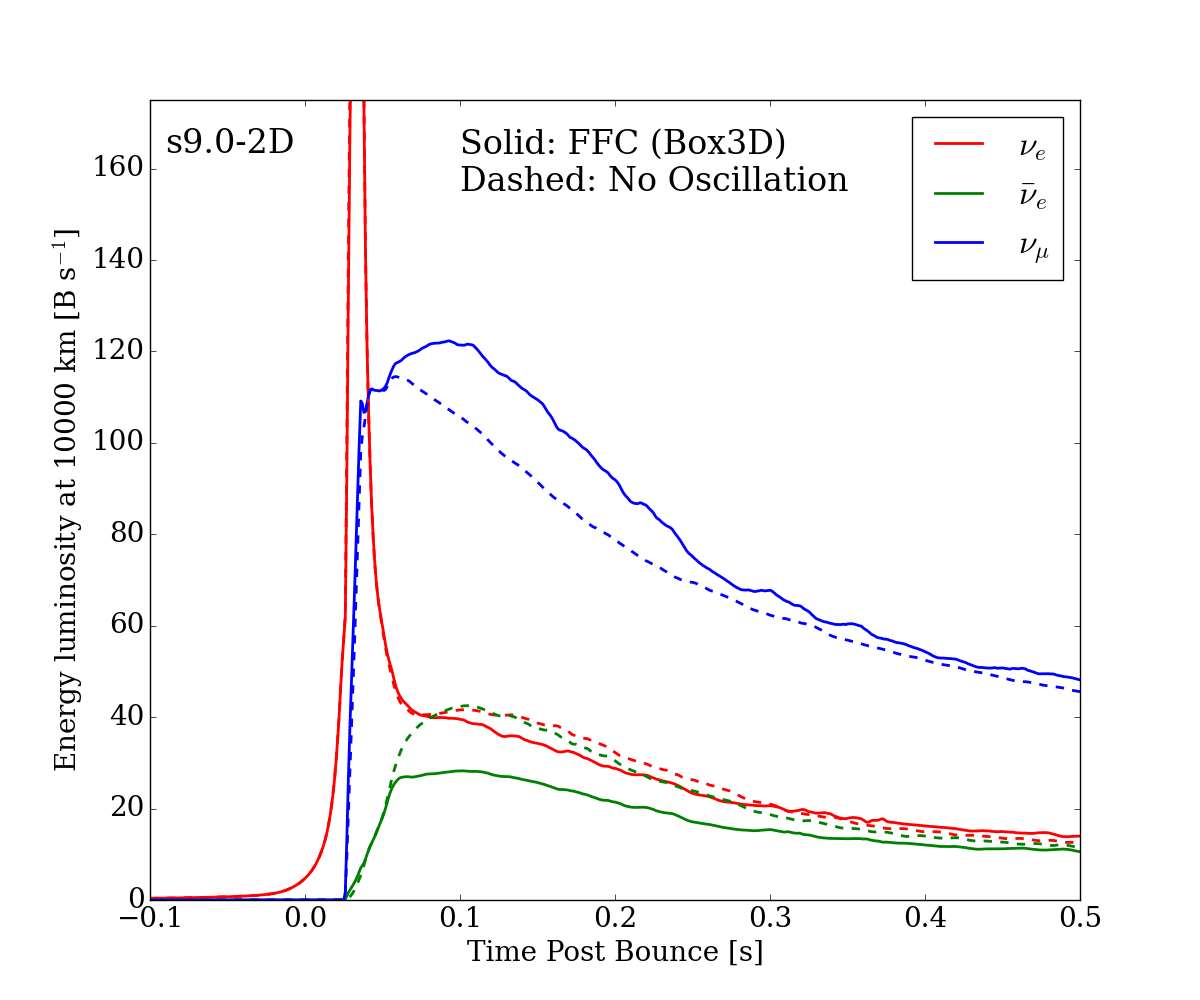}
    \includegraphics[width=0.48\textwidth]{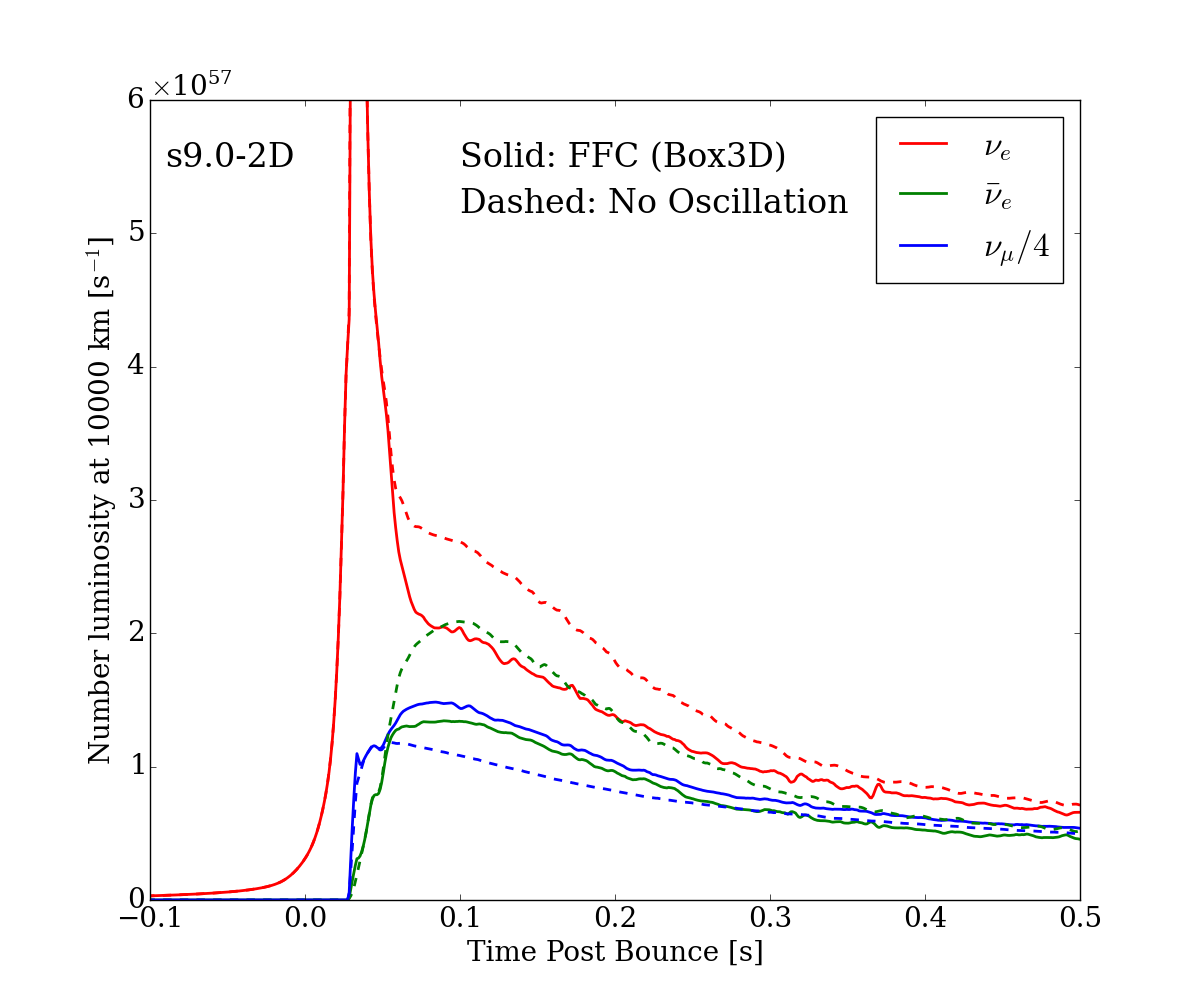}    
    \includegraphics[width=0.48\textwidth]{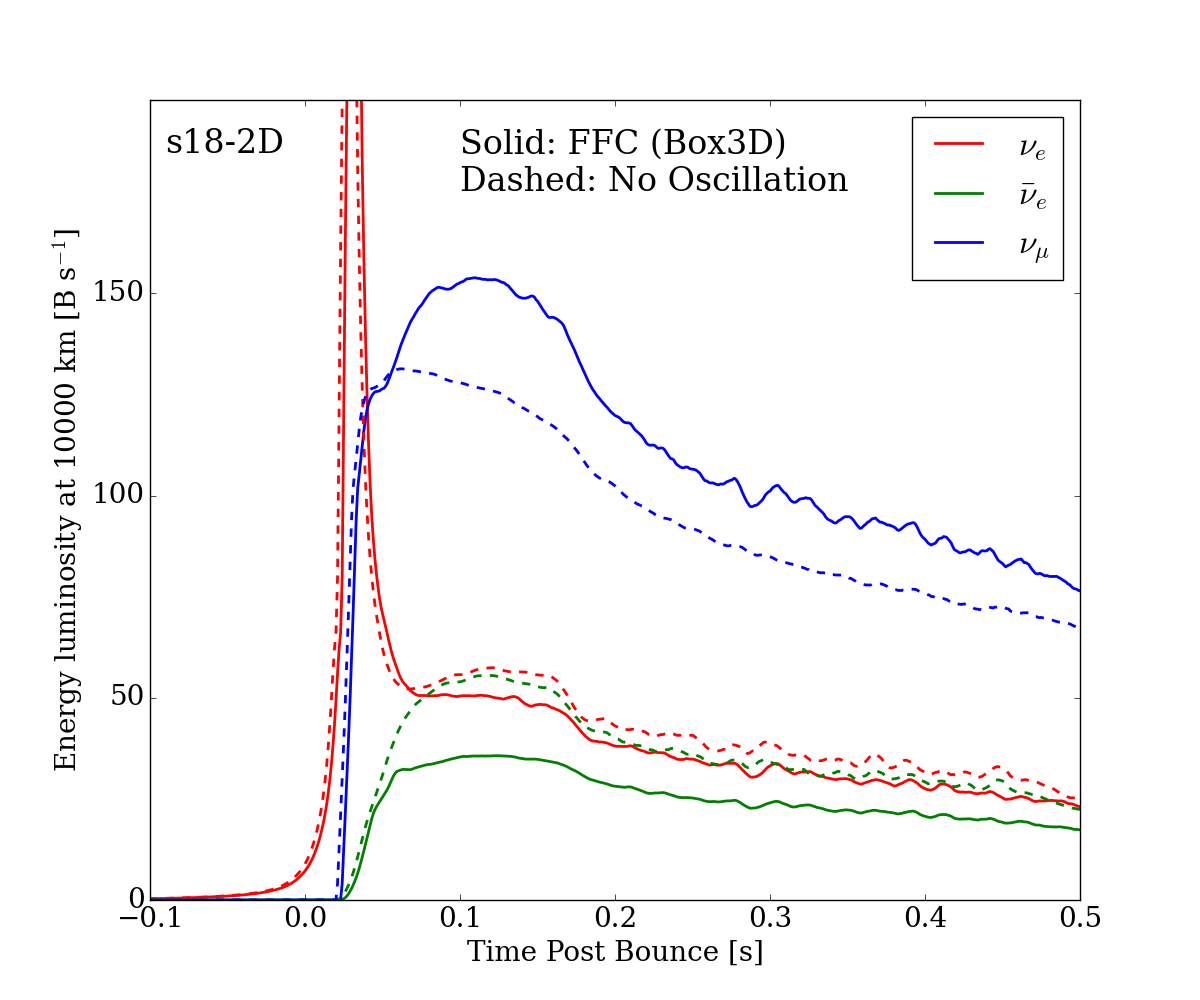}
    \includegraphics[width=0.48\textwidth]{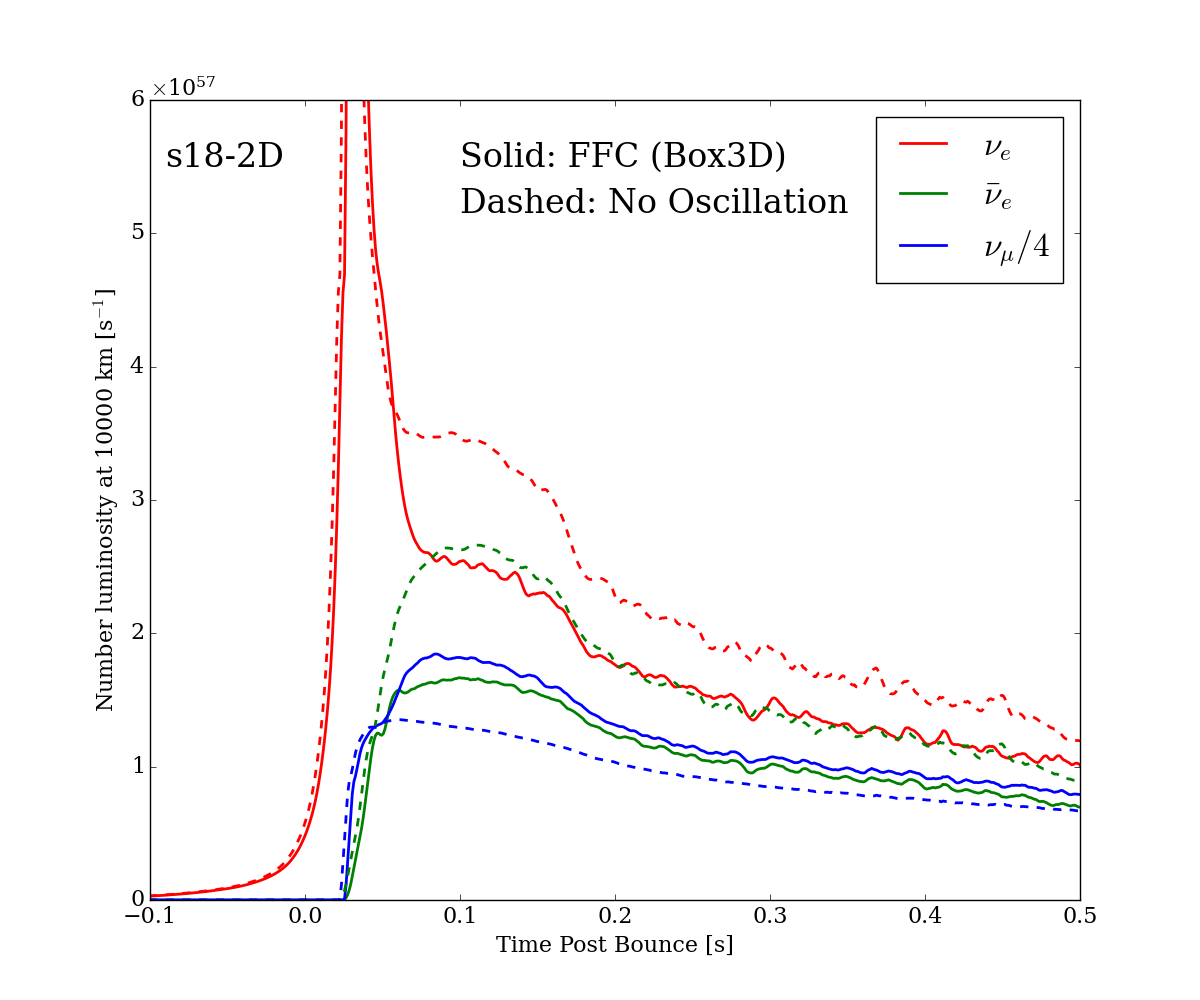}
    \caption{{\bf Left:} The evolution of the neutrino energy luminosities ($L$, in units of Bethes ($10^{51}$ erg) s$^{-1}$ at 10000 km) versus time after bounce (in seconds) for the 9 (top) and 18 (bottom) $M_{\odot}$ progenitor 2D simulations of Figure \ref{fig:heating-2d}, with (solid) and without (dashed) the inclusion of the FFC. {Note that $\nu_\mu$ is the combination of $\nu_\mu$, $\bar{\nu}_\mu$, $\nu_\tau$, and $\bar{\nu}_\tau$.} The behaviors are very similar to the corresponding 1D simulations, except that 2D models have higher luminosities at later time than 1D because of proto-neutron star convection \citep{Nagakura2020}. {\bf Right:} The corresponding number luminosities (in units of $10^{57}$ neutrinos per second at 10000 km) versus time after bounce. {For the number luminosity comparison, we show $``\nu_\mu"/4$.} While the FFC does redistribute the neutrino species mix, equipartition is not fully achieved in these simulations on our finite grid. See text for a discussion.}
    \label{fig:luminosities-2d}
\end{figure*}

\begin{figure*}
    \centering
    \includegraphics[width=0.48\textwidth]{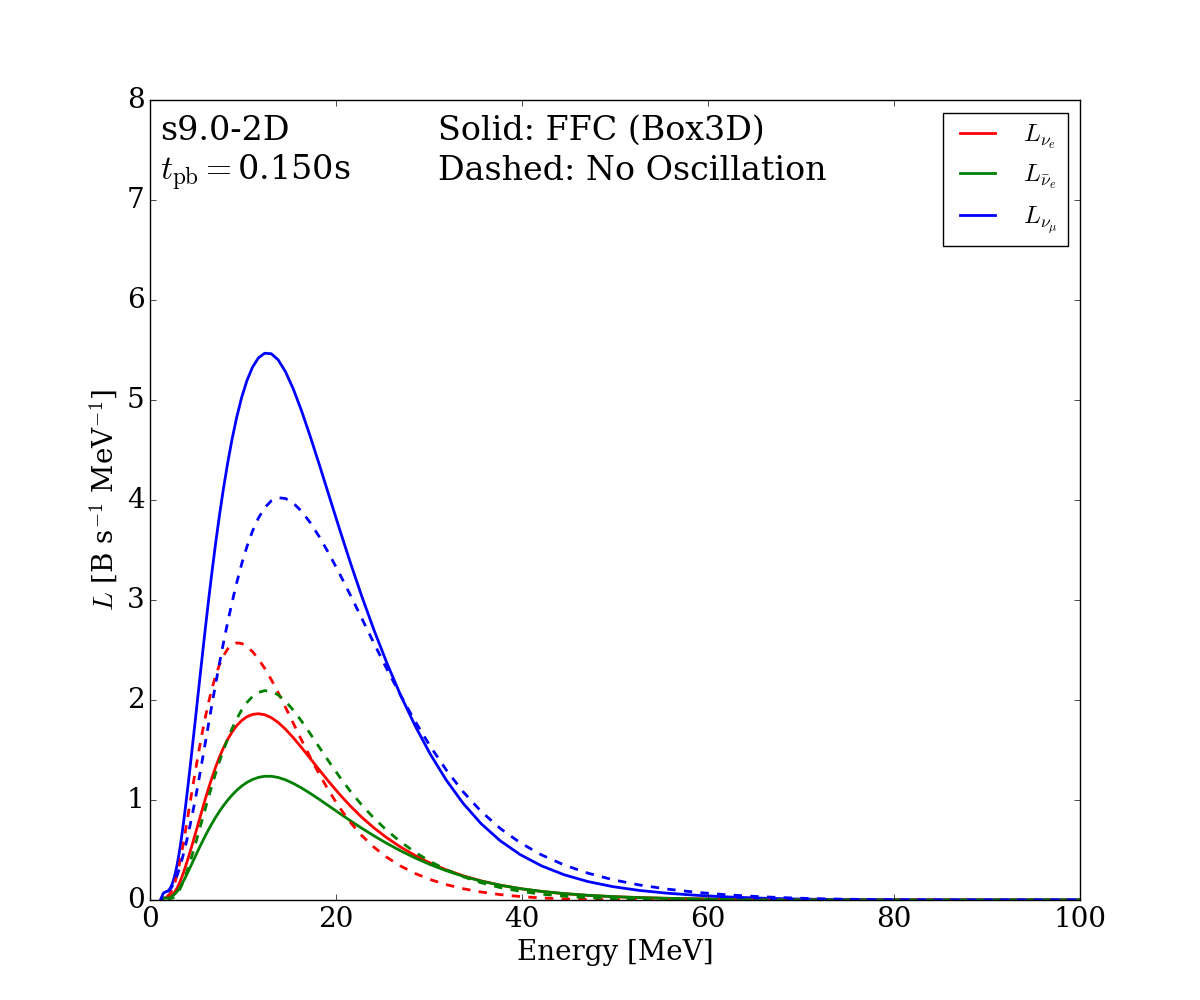}
    \includegraphics[width=0.48\textwidth]{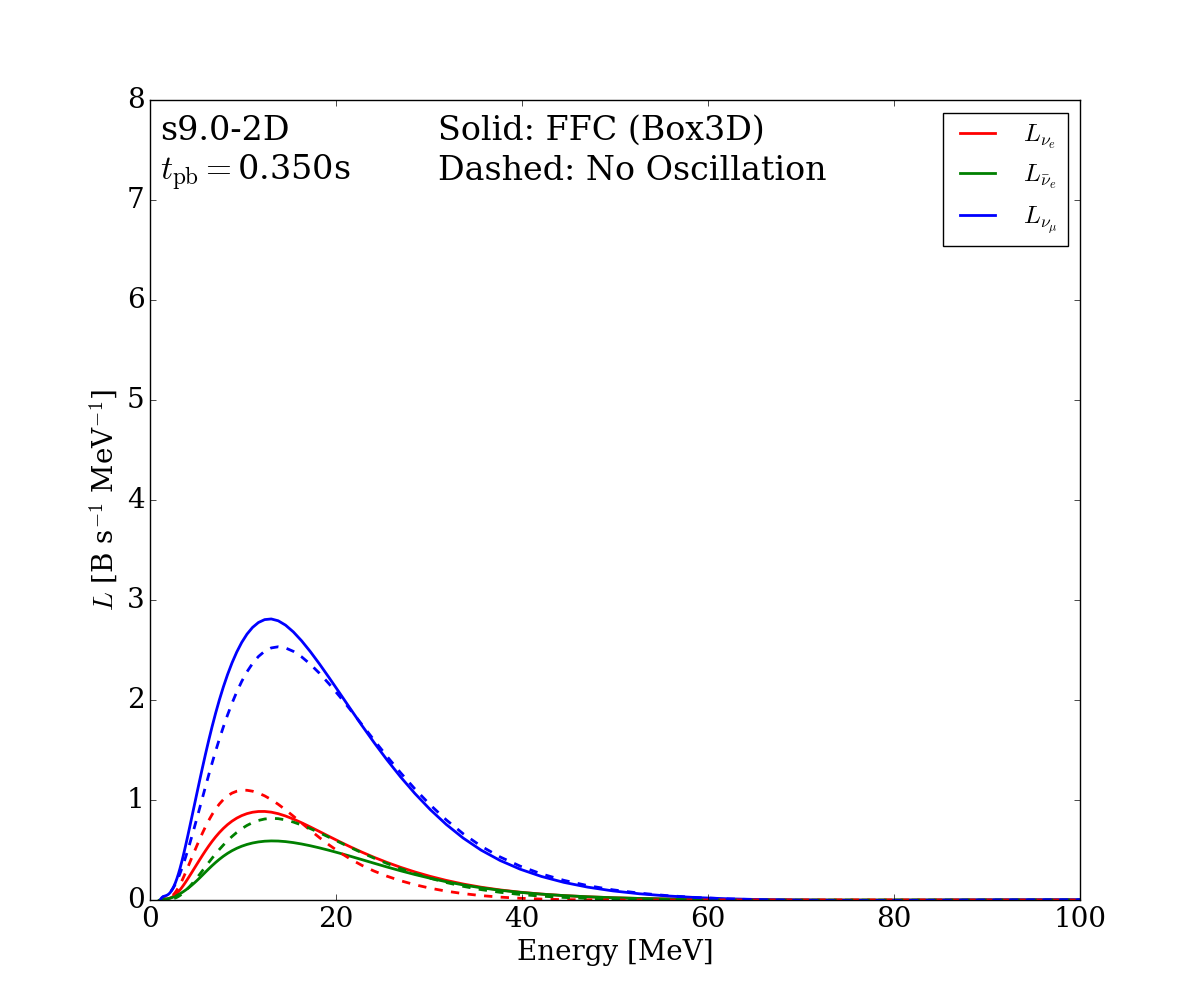}
    \includegraphics[width=0.48\textwidth]{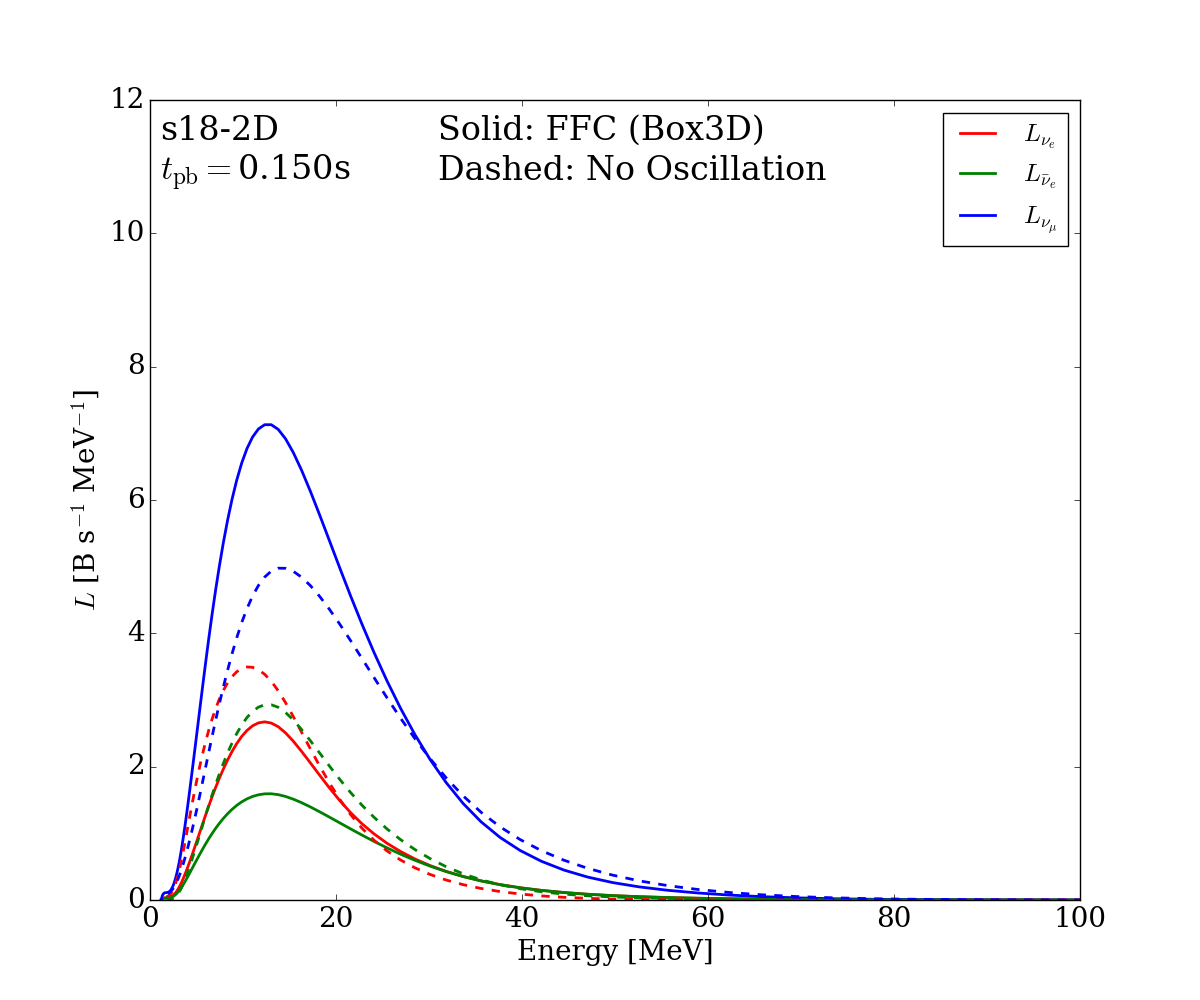}
    \includegraphics[width=0.48\textwidth]{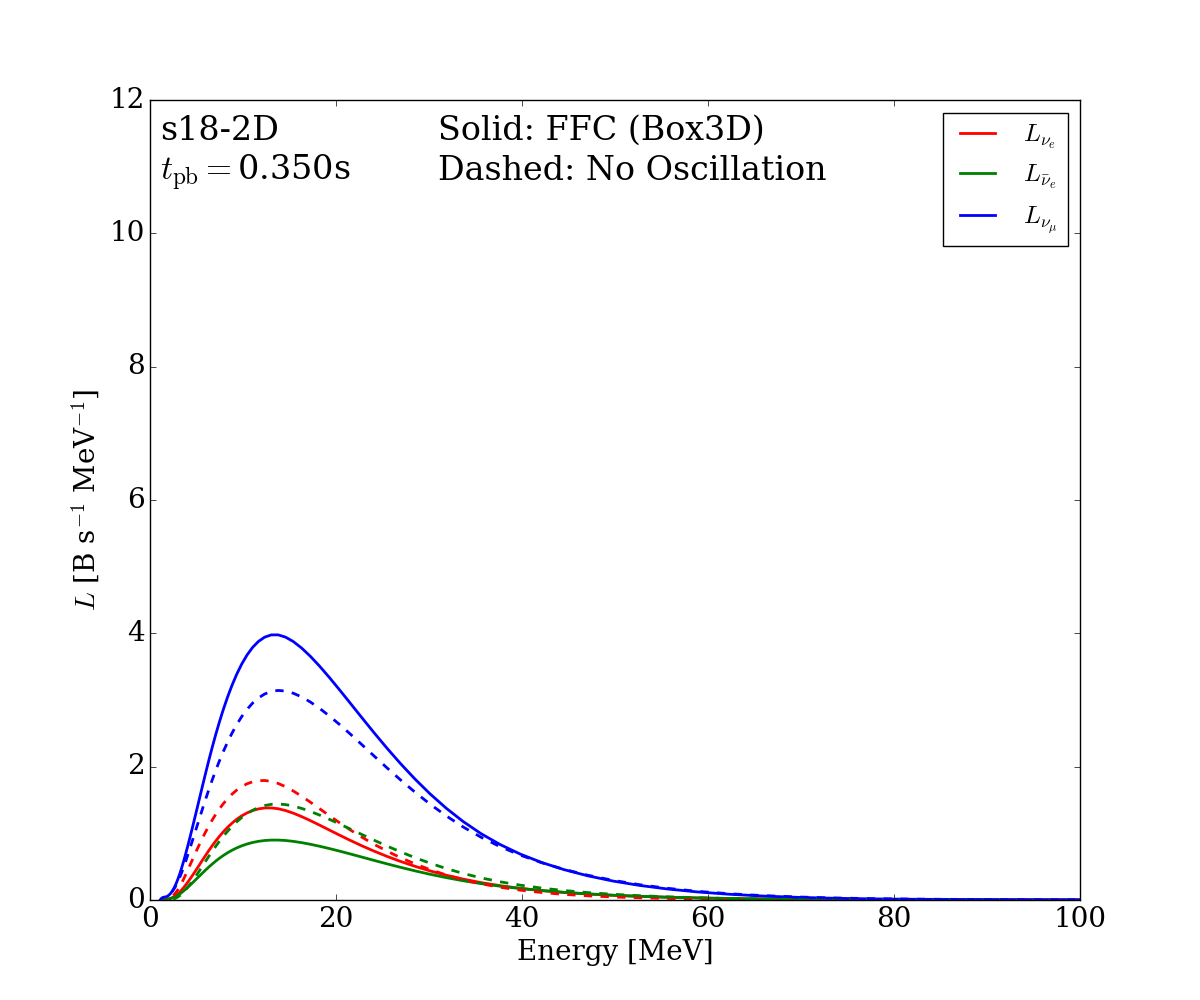}
    \caption{{The emergent spectra (splined) of the neutrinos, with (solid) and without (dashed) the FFC, for 2D simulations of the progenitors highlighted in Figure \ref{fig:luminosities-2d} and for two times after bounce. {Note that $\nu_\mu$ is the combination of $\nu_\mu$, $\bar{\nu}_\mu$, $\nu_\tau$, and $\bar{\nu}_\tau$.} When the FFC is operative, the $\nu_{\mu}$ neutrino spectrum is softened, while both the $\nu_e$ and $\bar{\nu}_e$ neutrino spectra harden slightly. The net effect on the neutrino-matter heating rate is a slight enhancement (see Figures \ref{fig:heating} and \ref{fig:heating-2d}).}} 
    \label{fig:spectra-2d}
\end{figure*}

\begin{figure*}
    \centering
    \includegraphics[width=0.48\textwidth]{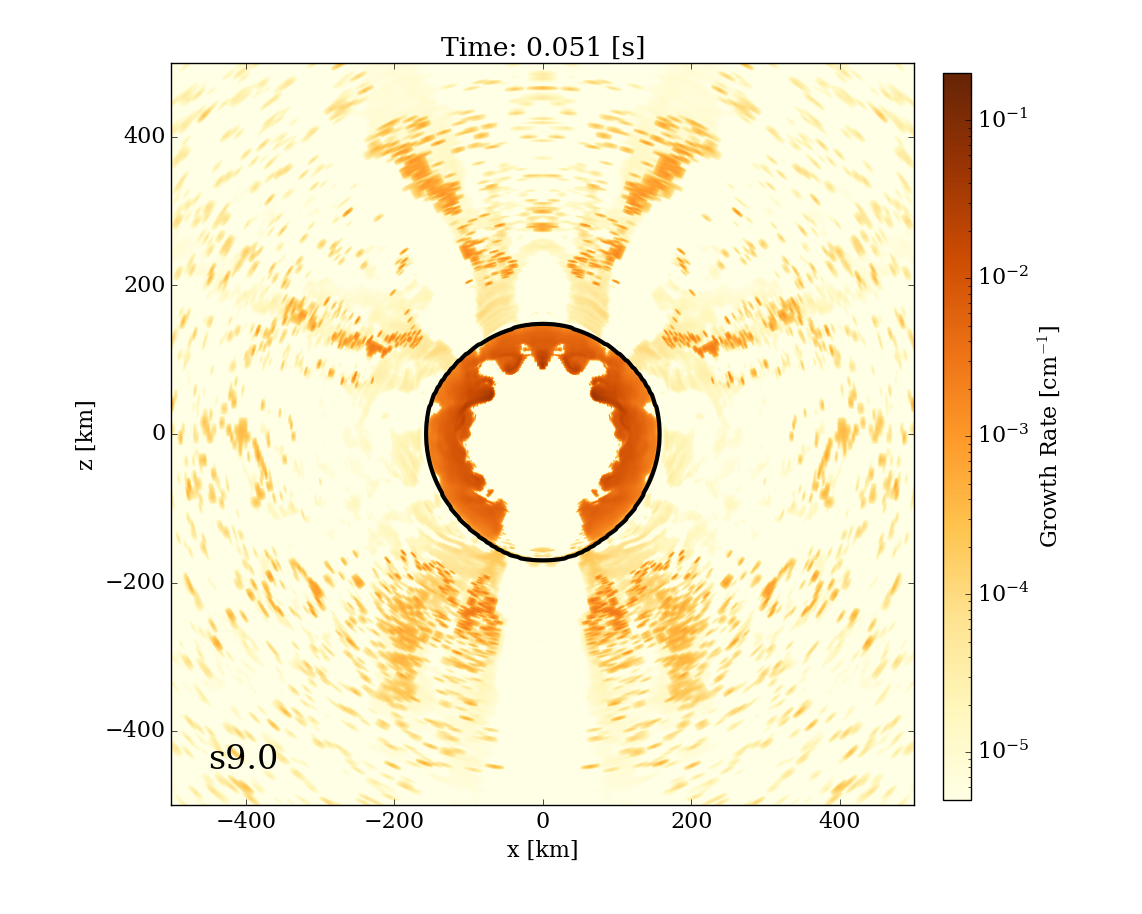}
    \includegraphics[width=0.48\textwidth]{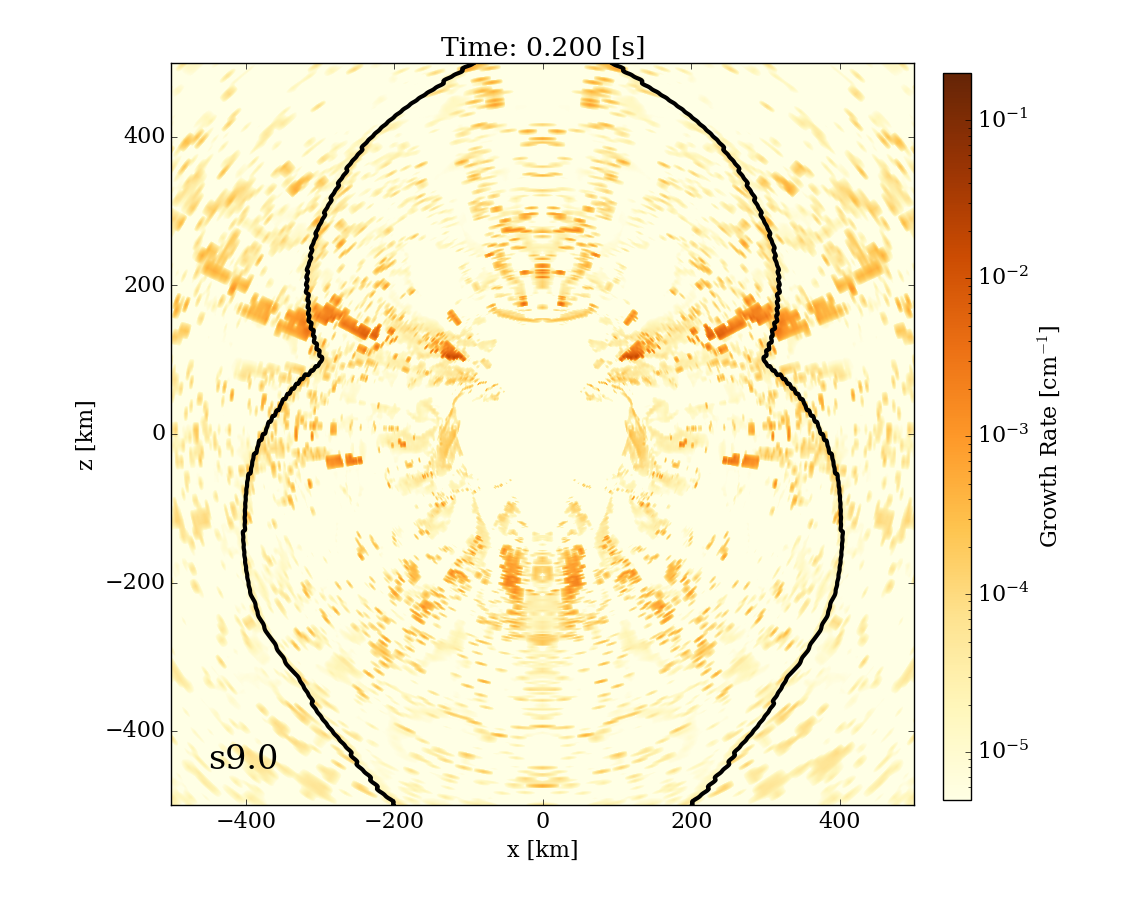}
    \includegraphics[width=0.48\textwidth]{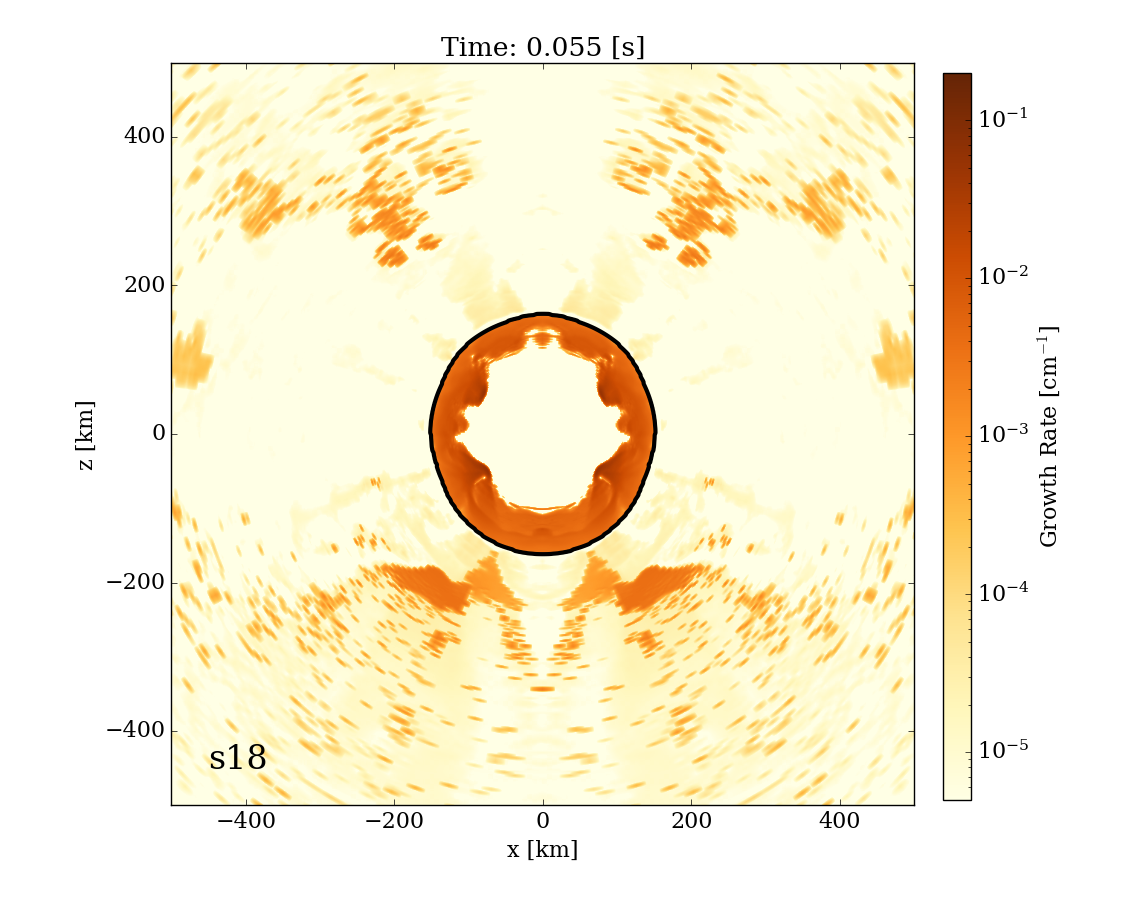}
    \includegraphics[width=0.48\textwidth]{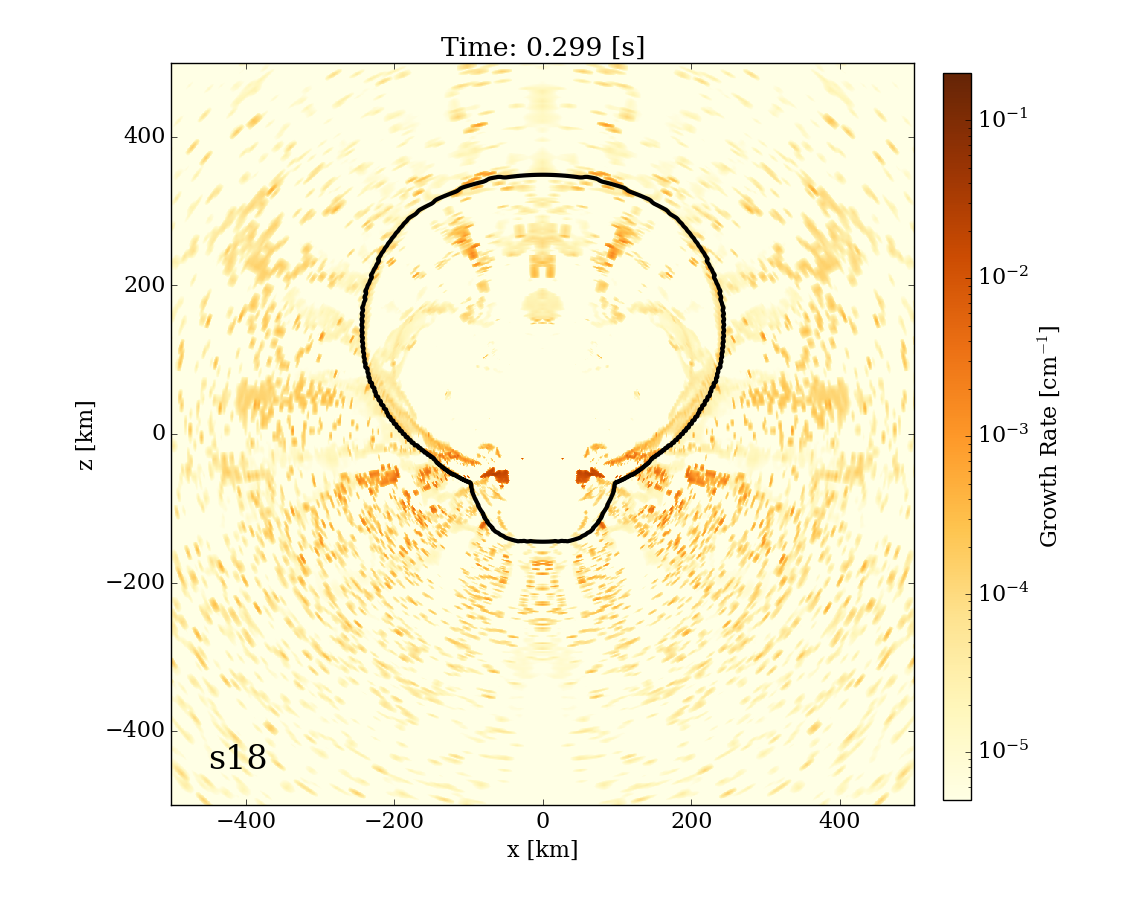}
    \caption{{The local {FFC growth} rate (using eq. \ref{eq:growth}, in units of cm$^{-1}$) for representative times after bounce for the 2D {9 and 18 $M_\odot$} models. The black contour shows the position of the shock. At early phases, there are {FFI} regions both interior and exterior to the shock, while at later phases the inner {FFI} region shrinks and disappears. Note the spike-like structures that are artifacts of the rings seen in 2D hydrodynamics that don't exist in 3D.}}
    \label{fig:ffc_region_2d}
\end{figure*}

\begin{figure*}
    \centering
    \includegraphics[width=0.48\textwidth]{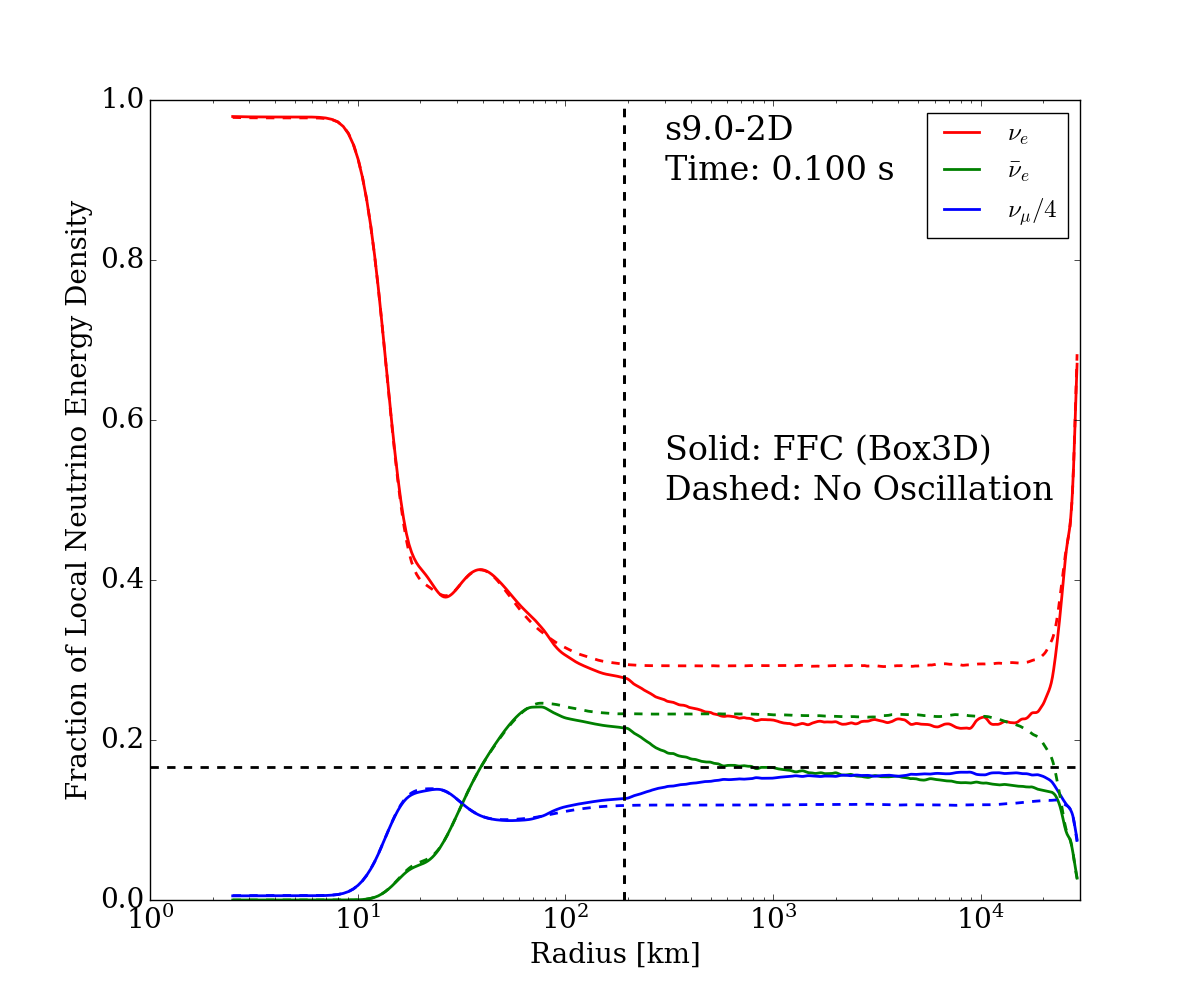}
    \includegraphics[width=0.48\textwidth]{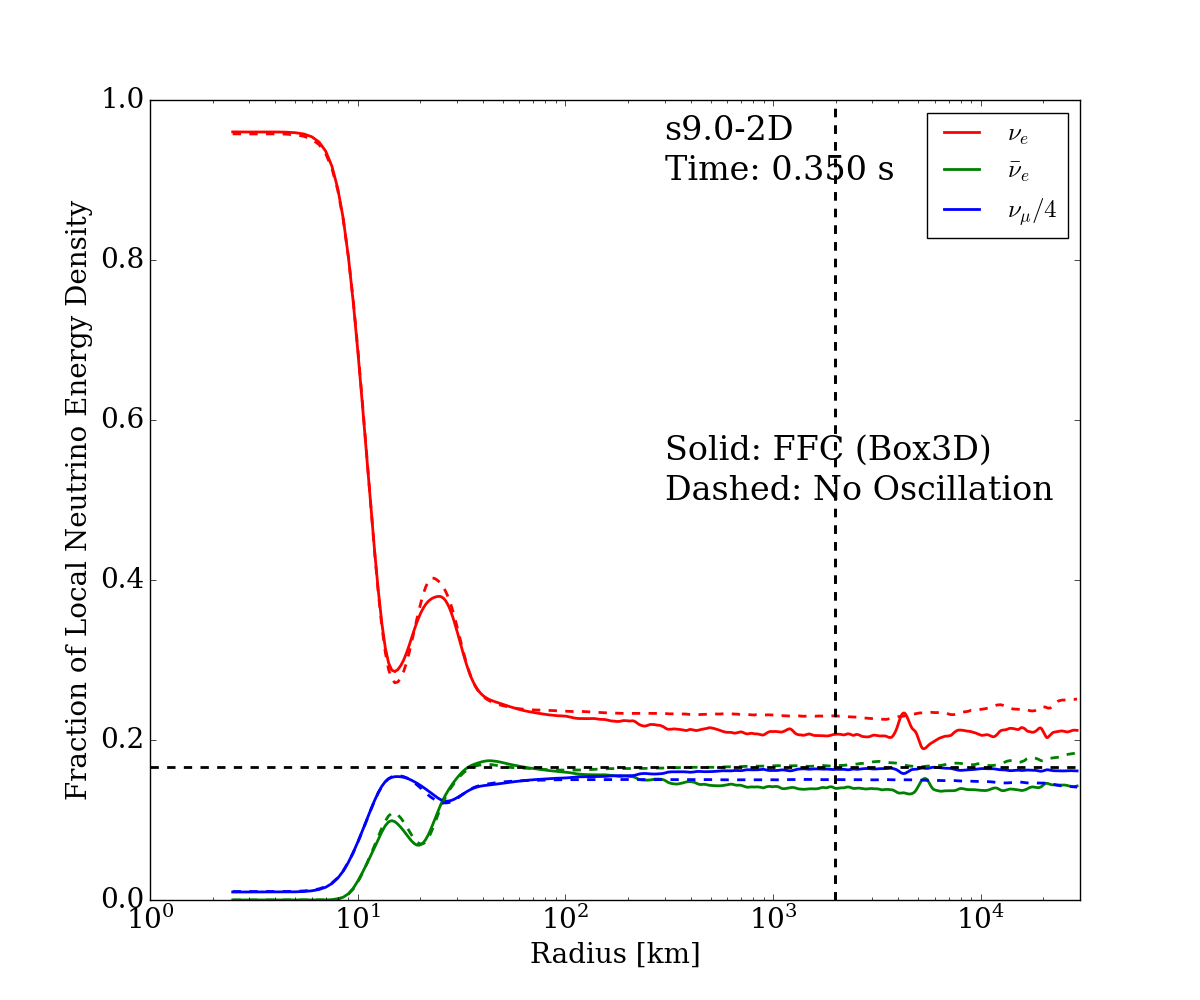}    
    \includegraphics[width=0.48\textwidth]{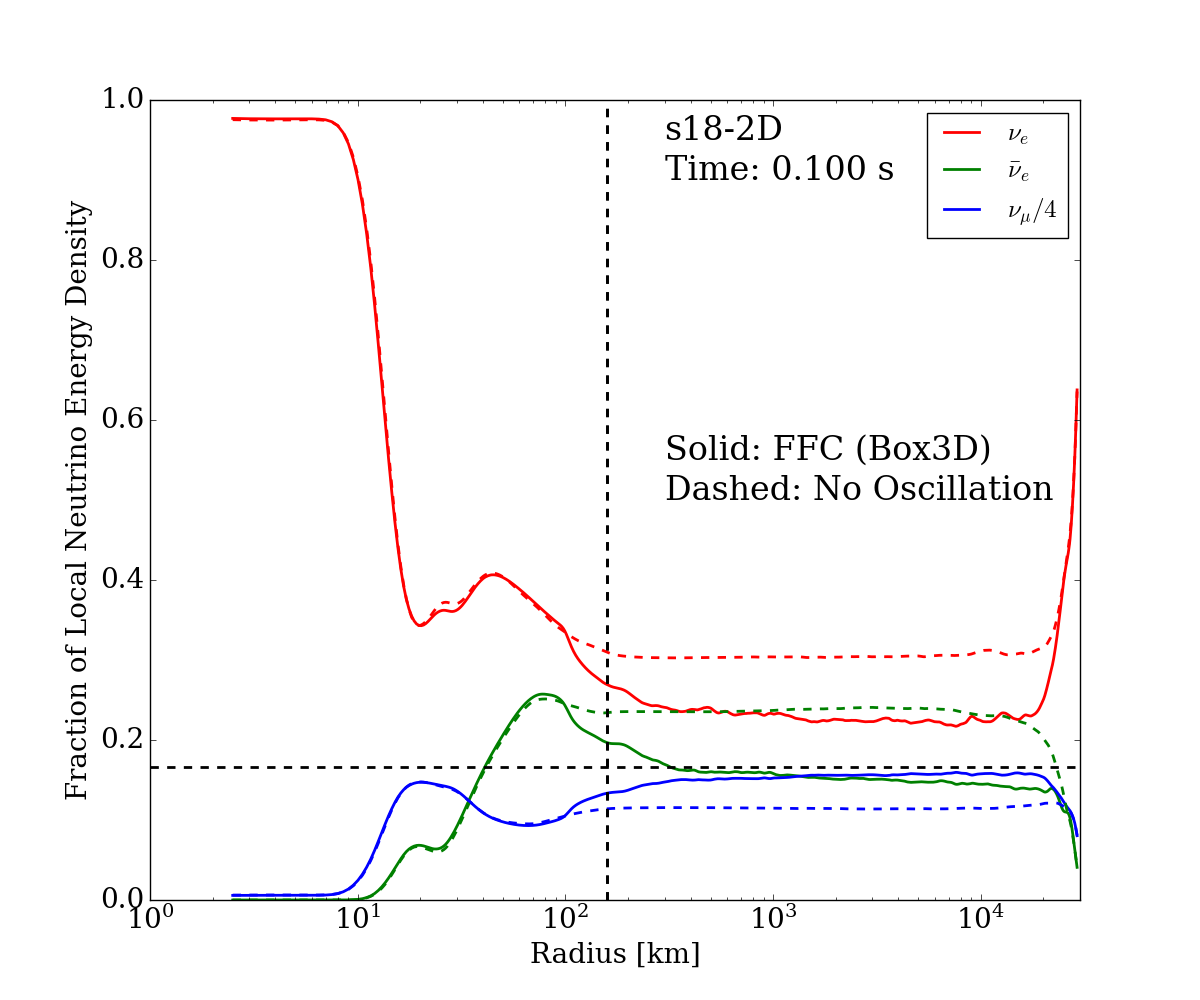}
    \includegraphics[width=0.48\textwidth]{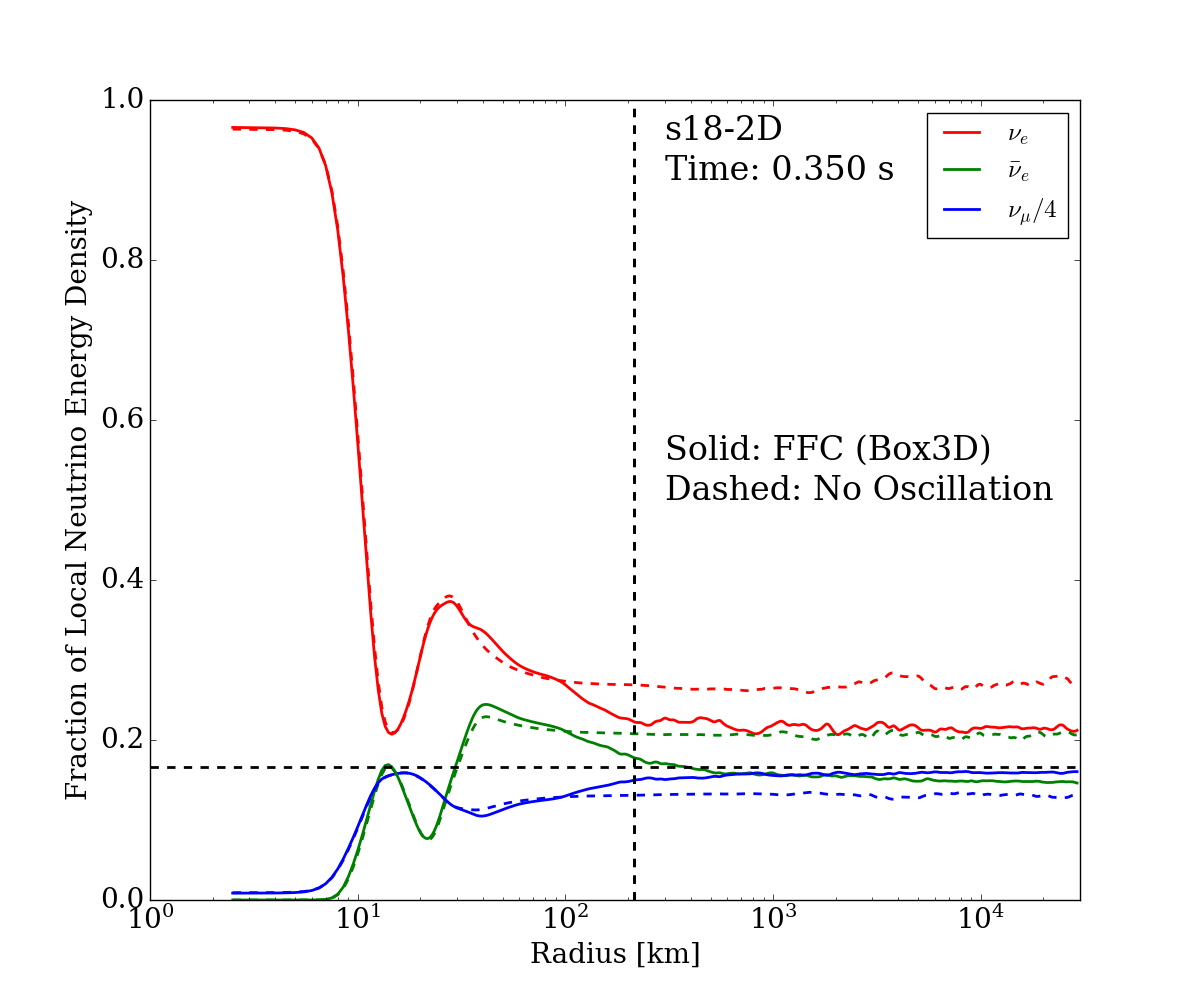}
    \caption{{The angle-averaged neutrino energy density fraction profiles for the 9 (top) and 18 (bottom) $M_{\odot}$ progenitor 2D simulations at two times after bounce, with (solid) and without (dashed) the inclusion of the FFC. Note that $\nu_\mu$ is the combination of $\nu_\mu$, $\bar{\nu}_\mu$, $\nu_\tau$, and $\bar{\nu}_\tau$ neutrinos. We have smoothed these curves with boxcar averages in radius (over 10 km). The vertical dashed line shows the averaged shock radius, while the horizontal dashed line at $1/6\approx0.167$ marks the equipartition fraction. The behavior is very similar to that shown in Figure \ref{fig:fractions}, except the 2D models explode.}}

    \label{fig:fractions-2d}
\end{figure*}

\clearpage

\bibliographystyle{aasjournal}
\bibliography{References}

\label{lastpage}
\end{document}